\documentclass[aps, prl, 10pt, notitlepage, nofootinbib, onecolumn, letterpaper, preprintnumbers, longbibliography, floatfix]{revtex4-1}
\pdfoutput=1 
\usepackage{amsmath,latexsym,amssymb,graphicx,slashed,hyperref,color,enumerate,url,etoolbox,cancel}
\hypersetup{colorlinks,citecolor= nicegreen,linkcolor= nicered}
\definecolor{nicered}{rgb}{0.7,0.1,0.1}
\definecolor{nicegreen}{rgb}{0.1,0.5,0.1}
\begin{document}
\def\Carleton{Department of Physics, Carleton University, Ottawa, Ontario K1S 5B6, Canada}

\title{On Dark Matter Self-interaction via Single Neutrino Exchange Potential}
\author{Yue Zhang}
\affiliation{\Carleton}
\date{\today}
\begin{abstract}
Neutrinos -- amongst the lightest known particles -- can mediate a force driving dark matter self-interaction and the small scale structure of the universe. We explore such a possibility in the simplest neutrino portal dark sector model where neutrino has a Yukawa coupling with a scalar $\phi$ and fermion $\chi$ that are degenerate in mass and together comprise 100\% of dark matter in the universe. We derive the non-relativistic potential generated by single-neutrino exchange which is of the monopole-dipole form and explore $\chi\phi\to\chi\phi$ scattering based on phase shift analysis. Our result shows that Born approximation continues to be valid in the low energy regime and the scattering cross section scales as $1/v^2$ over a wide range of dark matter velocities. Such a velocity-dependent self-interacting cross section can be large enough to explain the shallow density of dwarf galaxy cores and consistent with the upper limit from colliding galaxy clusters. The $1/v^2$ behavior persists down to rather low velocities $v\sim m_\nu/m$ where $m$ is the dark matter mass, leaving the opportunity for further astrophysical probes. Through the neutrino portal, the self-interacting dark matter parameter space can be tested by searches for $Z$-boson and light mesons decaying into the dark sector, as well as low-mass dark matter direct detection experiments.
\end{abstract}

\maketitle

Dark matter has been known to exist throughout the universe but its fundamental nature remains mysterious. Many dark matter candidates have been suggested in the literature based on different principles,
along with the opportunity to uncover their identities experimentally. 
Although a broad search program is being developed and some dark matter candidates are under thorough investigation, a discovery is still lacking and not guaranteed.
Without further observational evidence, it is not clear which points to the correct path forward.

The self-interacting dark matter was proposed based on potential clues in small-scale structure observations~\cite{Spergel:1999mh, Dave:2000ar}. With a cross section per dark matter mass in the range of $(0.1-10)\,{\rm cm^2} /{\rm gram}$, recent analyses demonstrate that the self-interaction can modify the dark matter mass density distribution by lowering the core density of dwarf galaxies and improve the agreement with observations, while retaining the successful predictions of collisionless cold dark matter in large scale structure (see~\cite{Tulin:2017ara, Adhikari:2022sbh} for reviews).
It serves as a leading alternative to the baryonic feedback mechanism and provides an appealing motivation to explore the underlying theories.

Candidates for self-interacting dark matter can be found in various dark sector models~\cite{Feng:2009mn, Buckley:2009in,Loeb:2010gj,Aarssen:2012fx,Tulin:2012wi,Tulin:2013teo,Bellazzini:2013foa,Boddy:2014yra,Hochberg:2014kqa,Soni:2016gzf,Zhang:2016dck,Blennow:2016gde,Soni:2017nlm,McDermott:2017vyk,Chu:2018fzy,Chu:2018faw,Chu:2019awd,Costantino:2019ixl,Agrawal:2020lea,Tsai:2020vpi,Chaffey:2021tmj}. In most cases, a light dark force carrier is introduced to mediate the dark matter self-interaction. A light mediator can also be used to generate a velocity-dependent self-interacting cross section enhanced at lower velocities -- a feature more favored by observations~\cite{Kaplinghat:2015aga}. 

In a recent work, we pointed out the possibility that dark matter self-interaction can occur via the exchange of the Standard Model neutrinos~\cite{Orlofsky:2021mmy}. The unique feature is to get rid of the need for additional light mediator states. Instead, the role is fully played by the active neutrinos which are known particles and have small masses. This naturally occurs in a class of neutrino portal dark matter models which introduces Yukawa interaction of a neutrino with a fermionic dark matter and a scalar dark partner. After integrating out the heavier partner, the resulting effective Lagrangian is a four fermion interaction with two dark matter and two neutrino fields. 
Dark matter self interaction occurs via double-neutrino exchanges at one-loop level. The corresponding quantum mechanics potential has the $1/r^5$ form in the massless neutrino limit. The resulting self-interaction cross section is dominated by $S$-wave contribution and velocity independent.

In this note, we take a further step to explore dark matter self-interaction in the above framework in the limit where the dark fermion and scalar are degenerate in mass and both serve as dark matter that fills the universe. The fermion and scalar scattering effectively contributes to dark matter self-interaction and their potential is generated by the exchange of a single neutrino. We will derive the corresponding self-interaction cross section based on a partial wave analysis and show it has phenomenologically interesting velocity dependent behaviors.

\section{Model}

We begin with a gauge invariant Lagrangian that resembles neutrino portal dark sector models~\cite{Falkowski:2009yz, Ko:2014bka, Bertoni:2014mva, Batell:2017rol, Berryman:2017twh, Schmaltz:2017oov, Batell:2017cmf, Becker:2018rve, Folgado:2018qlv, Lamprea:2019qet, Zhang:2020nis}
\begin{equation}\label{eq:LAM}
\mathcal{L} = \frac{(\bar\chi \phi)(H^\dagger L)}{\Lambda} + {\rm h.c.} \ ,
\end{equation}
where $\chi$ is a gauge-siglet Dirac fermion and $\phi$ is a gauge-singlet complex scalar. For simplicity, we work in the limit that they have equal mass $m$ and both serve as the dark matter. 
$L$ is a Standard Model lepton doublet and $H$ is the Higgs doublet field. The above dimension five operator can be generated by integrating out a right-handed neutrino that connects the visible and dark sector. Below the electroweak symmetry breaking scale, we get
\begin{equation}\label{eq:Yukawa}
\mathcal{L} = y \bar \chi \mathbb{P}_L \nu \phi + {\rm h.c.} \ ,
\end{equation}
where $y = \langle H\rangle/\Lambda$ is a dimensionless Yukawa coupling and $\mathbb{P}_L=(1-\gamma_5)/2$. The masses of $\chi, \phi$ of interest to this work are well above neutrino mass scale.

Dark matter self-interaction in this model can occur through various processes because $\chi$ and $\phi$ are both dark matter. Tree-level processes with a virtual neutrino exchange include 
$\chi\phi\to\chi\phi$, $\chi\bar\chi\leftrightarrow \phi\phi^*$ and $\chi\phi^*\to\chi\phi^*$. Our following discussions will focus on the first process. If the dark matter relic density is made of particle-anti-particle asymmetries, the second and third processes can be turned off, so are their charge-conjugation channels.
We proceed with such a simplification by assuming the relic densities of $\chi$ and $\phi$ are both asymmetric. As will be discussed below, this assumption is consistent with the thermal freeze out scenario where antiparticles can be efficiently annihilated away.

\begin{figure}[h]
  \begin{center}
  \includegraphics[width=0.618\textwidth, height=5.2cm]{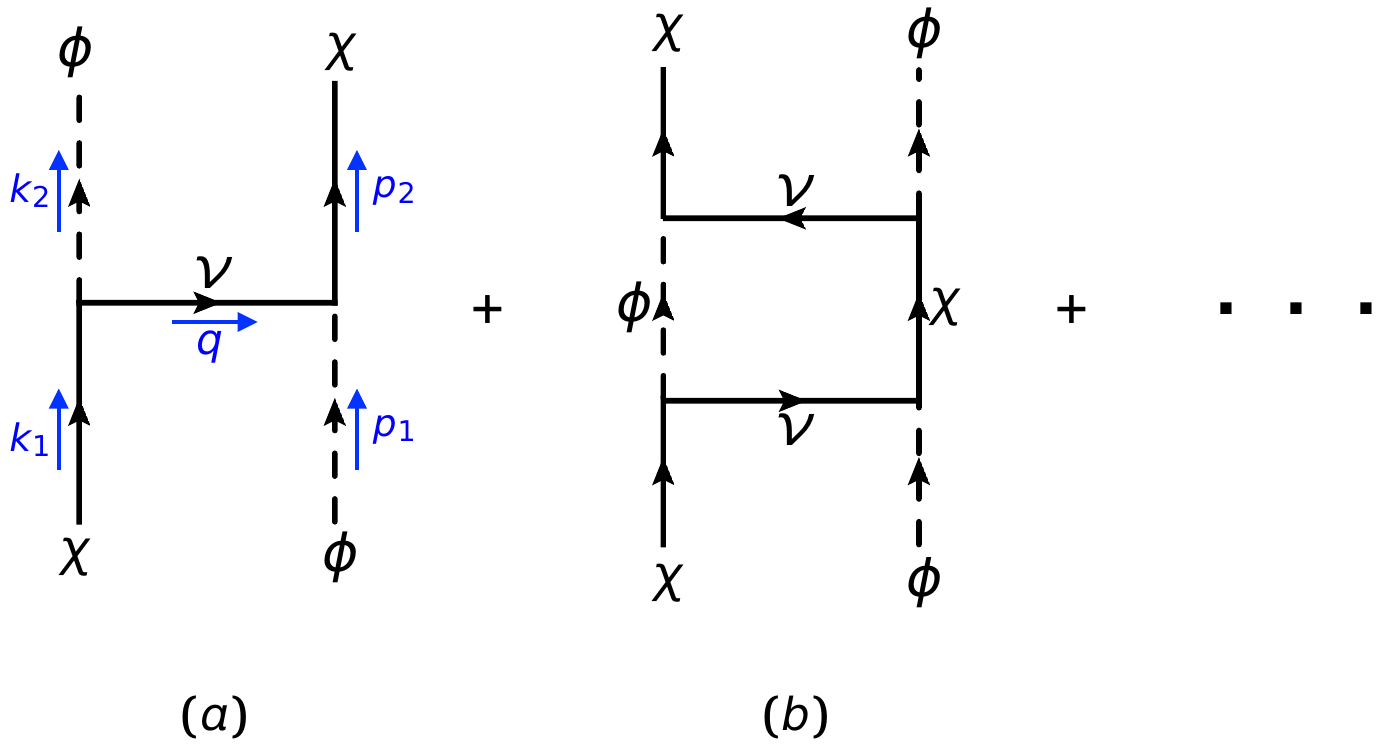}\vspace{-0.3cm}
  \end{center}
  \caption{Tree and loop level processes for $\chi,\phi$ scattering with neutrino exchanged in between. The are both dark matter. Time flows upwards. The arrows on the $\chi$ and $\phi$ lines indicate the flow of dark particle number under a global $U(1)_D$ symmetry, where they carry same charge.}
  \label{FIG1}
\end{figure}

The Feynman diagrams for the $\chi \phi \to \chi \phi$ scattering are shown in Fig.~\ref{FIG1} which also include the loop-level processes. The tree-level cross section can be easily computed in the relativistic theory. The total cross section in the low energy limit takes the form
\begin{equation}\label{eq:xsec0}
\sigma  = \frac{y^4}{64\pi m^2 v^2} \left[\log \left(\frac{m_\nu^2 + m^2 v^2}{m_\nu^2}\right) - \frac{m^2 v^2}{m_\nu^2 + m^2 v^2} \right] \ ,
\end{equation}
where $v\ll1$ is the relative velocity between $\chi$ and $\phi$. It is divergent in the $m_\nu=0$ limit.
More relevant for self-interacting dark matter are the momentum-transfer and viscosity cross sections,
\begin{equation}\label{eq:xsecTV}
\begin{split}
\sigma_T &\equiv \int (1-\cos\theta) d\cos\theta \frac{d\sigma}{d\cos\theta} = \frac{y^4}{32\pi m^2 v^2} \left[1 + \frac{m_\nu^2} {m^2 v^2} - \frac{m_\nu^4}{m^2v^2 (m_\nu^2 + m^2 v^2)} + \frac{2 m_\nu^2}{m^2 v^2} \log \left(\frac{m_\nu^2}{m_\nu^2 + m^2 v^2}\right)  \right] \ , \\
\sigma_V &\equiv \int (1-\cos^2\theta) d\cos\theta \frac{d\sigma}{d\cos\theta} = \frac{y^4}{32\pi m^2 v^2} \left[1 + \frac{6m_\nu^2} {m^2 v^2} + \frac{2m_\nu^2(3 m_\nu^2 + 2m^2 v^2)}{m^4 v^4} \log \left(\frac{m_\nu^2}{m_\nu^2 + m^2 v^2}\right)  \right] \ . \\
\end{split}
\end{equation}
They remain finite for any values of neutrino mass,
\begin{equation}\label{eq:xsec}
\sigma_T = \left\{ \begin{array}{cc}
\displaystyle{\frac{y^4}{32\pi m^2 v^2}}\ , &\quad m_\nu \ll mv \\[2ex]
\displaystyle{\frac{y^4 m^2 v^2}{96\pi m_\nu^4}} \ , &\quad m_\nu \gg mv
\end{array}
\right. \hspace{2cm}
\sigma_V = \left\{ \begin{array}{cc}
\displaystyle{\frac{y^4}{32\pi m^2 v^2}}\ , &\quad m_\nu \ll mv \\[2ex]
\displaystyle{\frac{y^4 m^2 v^2}{192\pi m_\nu^4}} \ , &\quad m_\nu \gg mv
\end{array}
\right.
\end{equation}

By only doing the Born-level calculation, it is not clear whether the above $\sigma_T, \sigma_V$ results would still hold when the scattering occurs at low energies, especially when $m_\nu/m \ll v\ll1$ that applies to a wide range of dark matter environments from dwarf galaxies to clusters. As a well-known example, for dark matter self-interaction mediated by a light dark photon/scalar, multiple dark photon/scalar exchanges are important and the ladder diagrams must be resummed if the scattering occurs outside the Born regime~\cite{Buckley:2009in, Tulin:2012wi, Tulin:2013teo}. For neutrino mediated dark force considered here, Fig.~\ref{FIG1} (b) shows the diagram with two neutrino exchanged in the t-channel~\footnote{Crossing the two neutrino propagators in Fig.~\ref{FIG1} (b) turns the $\chi \phi \to \chi \phi$ process into $\chi \phi^* \to \chi \phi^*$. The latter is irrelevant under the asymmetric dark matter assumption.} and $\dots$ represents the ladder diagrams that exchange more than two neutrinos (higher loops).
Rather than directly computing those loop diagrams, we will solve the non-relativistic Schr\"odinger equation for the $\chi$-$\phi$ scattering problem.

\section{Non-relativistic Interacting Potential}

To derive the leading-order interacting potential, we take the low-energy limit of the tree-level scattering amplitude using the non-relativistic reduction of scalar and fermion fields
\begin{equation}
\chi = e^{-imt} \sqrt{\frac{E+m}{2E}} \begin{pmatrix}
    \psi\\
    \frac{\vec{\sigma}\cdot\vec{p}}{E+m} \psi
\end{pmatrix} \ , \quad 
\phi = \frac{e^{-imt}}{\sqrt{2m}} \varphi \ ,
\end{equation}
where $\psi$ and $\varphi$ are the non-relativistic fermion and scalar fields. 
The anti-particle modes are left out for asymmetric dark matter considered here.
In the corresponding Dirac equation of $\chi$, the convention for $\gamma$ matrices is given in~\cite{Cheng:1984vwu}. Using the momentum assignment in Fig.~\ref{FIG1} (a), the matrix element for $\chi(k_1) \phi(p_1)\to \phi(k_2)\chi(p_2)$ takes the form
\begin{equation}
i \mathcal{M} = -i \frac{y^2}{4m} \frac{1}{q^2 - m_\nu^2} \langle p_2 | \bar \chi \cancel{q} (1+\gamma_5) \chi |k_1 \rangle \ ,
\end{equation}
where the momentum transfer $q$ satisfies $q^\mu \simeq (0, \vec{q})$ in the low energy limit and
\begin{equation}
\cancel{q} = \begin{pmatrix}
  0 & -\vec{\sigma}\cdot \vec{q} \\
  \vec{\sigma}\cdot \vec{q} & 0
\end{pmatrix} \ , \quad \cancel{q}\gamma_5 = \begin{pmatrix}
 -\vec{\sigma}\cdot \vec{q} & 0 \\
  0 & \vec{\sigma}\cdot \vec{q}
\end{pmatrix} \ .
\end{equation}
With $|\vec{k}_1|, |\vec{p}_2|\ll m$, the leading term from the matrix element is from the $\cancel{q}\gamma_5$ term,
\begin{equation}
i \mathcal{M} \simeq - i \frac{y^2}{4m} \frac{1}{|\vec{q}|^2 + m_\nu^2} \xi'^\dagger \vec{\sigma}\cdot \vec{q} \xi \ ,
\end{equation}
where $\xi, \xi'$ are the initial and final state Pauli spinors from the two component fermion field $\psi$.
The corresponding interacting potential in the momentum space is
\begin{equation}
V(\vec{q}) = \frac{y^2}{4m} \frac{\vec{\sigma}\cdot \vec{q}}{|\vec{q}|^2 + m_\nu^2} \ .
\end{equation}
Fourier transforming to coordinate space, we get
\begin{equation}\label{eq:v(r)}
\begin{split}
    V(\vec{r}) &= \int\frac{d^3\vec{q}}{(2\pi)^3} e^{i\vec{q}\cdot \vec{r}} V(\vec{q}) \\
    & = \frac{i y^2(1+m_\nu r) e^{-m_\nu r}}{16\pi m r^2} \vec{\sigma}\cdot \hat{r} \ ,
\end{split}
\end{equation}
where $r=|\vec{r}|$ and $\hat r$ is the unit vector along the $\vec{r}$ direction. This potential has the form of a monopole-dipole interaction~\cite{Dobrescu:2006au}.

The extra factor $i$ in $V(\vec{r})$ may appear puzzling because it make $V(\vec{r})$ anti-Hermitian, whereas the interacting potential as part of the Hamiltonian operator is expected to be Hermitian.
In fact, $V(\vec{r})$ should be considered as the off-diagonal interaction connecting two wavefunctions that describe the $\chi$-$\phi$ system, because it interchanges the spin between the two scattering particles. It is useful to consider a scattering state (denoted as $\Psi^+$) with the fermion $\psi$ traveling in the $+\hat z$ direction with momentum $\vec{p}$. With a very small momentum transfer ($|\vec{q}|\ll|\vec{p}|$) through the potential $V$ acting on the state, the momentum $\vec{p}$ remains almost the same but it is now carried by the scalar particle $\phi$ (see Fig.~\ref{FIG1} (a)). Meanwhile, the momentum of the fermion particle $\chi$ becomes approximately $-\vec {p}$. This is a different state from $\Psi^+$ which we will denote as $\Psi^-$.
The overall potential spanning in the $(\Psi^+, \Psi^-)$ space
\begin{equation}
\mathcal{V}(\vec{r}) = \begin{pmatrix}
0 & V(\vec{r}) \\
-V(\vec{r}) & 0
\end{pmatrix} \ ,
\end{equation}
is clearly Hermitian. The extra minus sign in the lower-left component arises because when repeating the above potential derivation using Fig.~\ref{FIG1} (a) but with the momenta of initial (and final) state $\chi$ and $\phi$ interchanged, the momentum transfer $q$ would go against the direction of arrow on the neutrino propagator.

The resulting time-independent Schr\"odinger equation takes the matrix form
\begin{equation}\label{eq:Schrodinger}
\left( - \frac{\vec{\nabla}^2}{2\mu} + \mathcal{V}({\vec{r}})\right) \begin{pmatrix}
\Psi^+_k(\vec{r}) \\\Psi^-_k(\vec{r})
\end{pmatrix} = \frac{k^2}{2\mu} \begin{pmatrix}
\Psi^+_k(\vec{r}) \\\Psi^-_k(\vec{r})
\end{pmatrix} \ ,
\end{equation}
where $\mu = m/2$ is the reduced mass and the $\Psi_k^\pm(\vec{r})$ here are stationary states. 
The momentum $k=\frac{1}{2}mv$ where $v$ is the relative velocity between $\chi$ and $\phi$, and
$k^2/(2m)$ is the kinetic energy carried by the $\chi$ or $\phi$ particle at $r\to \infty$ in the center of mass frame.

Throughout this work, we will make a realistic simplification by neglecting the neutrino mass in the above potential. A nonzero neutrino mass only strongly affect $V(\vec{r})$ at large distances $r > m_\nu^{-1}$ where the potential is smaller than $y^2 m_\nu^2/(16\pi m)$. It is negligible for a scattering problem where the dark matter velocity at infinity is much larger than $y m_\nu/(\sqrt{8\pi}m)$. The range of dark matter velocity and mass will be quantified in the dark matter phenomenology section. In the massless neutrino limit, the potential $V(\vec{r})$ can be written as
\begin{equation}
V(\vec{r}) = \frac{iy^2}{16\pi m}\frac{\vec{\sigma}\cdot \hat{r}}{r^2} = \frac{iy^2}{16\pi m r^2} \left[ \sqrt\frac{4\pi}{3} Y_{1,0}(\theta, \phi) \sigma_z + \sqrt\frac{8\pi}{3} Y_{1,-1}(\theta, \phi) \sigma_+ - \sqrt\frac{8\pi}{3} Y_{1,1}(\theta, \phi) \sigma_- \right] \ , 
\end{equation}
where where $Y_{\ell,m}(\theta, \phi)$ are the spherical harmonic functions, and
\begin{equation}
\begin{split}
\sigma_z =\begin{pmatrix}
    1 & 0 \\
    0 & -1
\end{pmatrix} \ , \quad \sigma_+ =\begin{pmatrix}
    0 & 1 \\
    0 & 0
\end{pmatrix} \ , \quad \sigma_- =\begin{pmatrix}
    0 & 0 \\
    1 & 0
\end{pmatrix} \ .
\end{split}
\end{equation}

\section{Partial Wave Expansion}

The interacting potential Eq.~\eqref{eq:v(r)} does not commute with the orbital angular momentum operator $\vec{L}$, $|\vec{L}|^2$ or the fermion spin operator $\vec{S}$, but rather their sum, the total angular momentum $\vec{J}=\vec{L}+\vec{S}$ and $|\vec{S}|^2$. In the following partial wave analysis we will decompose the total scattering wavefuntion in terms of common eigenstates of $\vec{J}^2$ and $\hat J_z$, with the corresponding eigenvalue $j(j+1)$ and $m_j$. Because the fermion $\chi$ has spin $S=\frac{1}{2}$, the total angular momentum $j$ can be made of states with orbital angular momentum $\ell=j\pm\frac{1}{2}$. 

Without loss of generality, we will work with the $m_j=\frac{1}{2}$ state hereafter and write the wavefunctions $\Psi^\pm_k(\vec{r})$ as
\begin{equation}\label{eq:Psi}
\Psi^\pm_k(\vec{r}) = \sum_{l=0}^\infty \left[ R^\pm_{\ell, \ell+\frac{1}{2},k}(r) \textstyle{\langle\theta,\phi\left|\ell, \frac{1}{2}, \ell+\frac{1}{2}, \frac{1}{2}\right\rangle} +
R^\pm_{\ell+1, \ell+\frac{1}{2},k}(r) \textstyle{\langle\theta,\phi\left|\ell+1, \frac{1}{2}, \ell+\frac{1}{2}, \frac{1}{2}\right\rangle}\right] \ ,
\end{equation}
where $|\ell sjm_j\rangle=\textstyle{|\ell, \frac{1}{2}, \ell\pm\frac{1}{2}, \frac{1}{2}\rangle}$ are linear combinations of $(\vec{L}^2, \hat L_z)$ and $(\vec{S}^2, \hat S_z)$ eigenstates via the Clebsch-Gorddn coefficients
\begin{equation}\label{eq:CB}
\begin{split}
  \textstyle{\langle\theta,\phi\left|\ell, \frac{1}{2}, \ell+\frac{1}{2}, \frac{1}{2}\right\rangle} &= \sqrt\frac{\ell+1}{2\ell+1} Y_{\ell,0}(\theta, \phi) {\textstyle\left|\frac{1}{2},\frac{1}{2}\right\rangle} + \sqrt\frac{\ell}{2\ell+1}Y_{\ell,1}(\theta, \phi) \textstyle{\left|\frac{1}{2},-\frac{1}{2}\right\rangle} \ ,
  \\
\textstyle{\langle\theta,\phi\left|\ell+1, \frac{1}{2}, \ell+\frac{1}{2}, \frac{1}{2}\right\rangle} &=
\sqrt\frac{\ell+1}{2\ell+3}Y_{\ell+1,0}(\theta, \phi) {\textstyle\left|\frac{1}{2},\frac{1}{2}\right\rangle} - \sqrt\frac{\ell+2}{2\ell+3}Y_{\ell+1,1}(\theta, \phi) \textstyle{\left|\frac{1}{2},-\frac{1}{2}\right\rangle}
\ .
\end{split}
\end{equation}

Acting $V(\vec{r})$ on $\Psi^\pm_k(\vec{r})$ and using the multiplication properties of spherical harmonics, after a lengthy but straightforward derivation we obtain 
\begin{equation}\label{eq:VPsi}
V(\vec{r}) \Psi^\pm_k(\vec{r}) = \frac{iy^2}{16\pi m r^2}\sum_{l=0}^\infty \left[ R^\pm_{\ell+1, \ell+\frac{1}{2},k}(r) \textstyle{\langle\theta,\phi\left|\ell, \frac{1}{2}, \ell+\frac{1}{2}, \frac{1}{2}\right\rangle} +
R^\pm_{\ell, \ell+\frac{1}{2},k}(r) \textstyle{\langle\theta,\phi\left|\ell+1, \frac{1}{2}, \ell+\frac{1}{2}, \frac{1}{2}\right\rangle}\right] \ .
\end{equation}

Plugging Eqs.~\eqref{eq:Psi} and \eqref{eq:VPsi} back into \eqref{eq:Schrodinger}, we obtain two sets of coupled equations for the radial wavefunctions $R^\pm_{\ell, \ell+\frac{1}{2},k}(r)$, $R^\pm_{\ell+1, \ell+\frac{1}{2},k}(r)$,
\begin{equation}\label{eq:SE1}
\begin{split}
\left[ - \frac{1}{r^2}\frac{\partial}{\partial r}r^2 \frac{\partial}{\partial r} + \frac{\ell(\ell+1)}{r^2} \right] R^+_{\ell, \ell+\frac{1}{2},k}(r) + \frac{iy^2}{16\pi r^2}R^-_{\ell+1, \ell+\frac{1}{2},k}(r) &= k^2 R^+_{\ell, \ell+\frac{1}{2},k}(r) \ , \\
\left[ - \frac{1}{r^2}\frac{\partial}{\partial r}r^2 \frac{\partial}{\partial r} + \frac{(\ell+1)(\ell+2)}{r^2} \right] R^-_{\ell+1, \ell+\frac{1}{2},k}(r) - \frac{iy^2}{16\pi r^2}R^+_{\ell, \ell+\frac{1}{2},k}(r) &= k^2 R^-_{\ell+1, \ell+\frac{1}{2},k}(r) \ ,
\end{split}
\end{equation}
and
\begin{equation}\label{eq:SE2}
\begin{split}
\left[ - \frac{1}{r^2}\frac{\partial}{\partial r}r^2 \frac{\partial}{\partial r} + \frac{\ell(\ell+1)}{r^2} \right] R^-_{\ell, \ell+\frac{1}{2},k}(r) - \frac{iy^2}{16\pi r^2}R^+_{\ell+1, \ell+\frac{1}{2},k}(r) &= k^2 R^-_{\ell, \ell+\frac{1}{2},k}(r) \ , \\
\left[ - \frac{1}{r^2}\frac{\partial}{\partial r}r^2 \frac{\partial}{\partial r} + \frac{(\ell+1)(\ell+2)}{r^2} \right] R^+_{\ell+1, \ell+\frac{1}{2},k}(r) + \frac{iy^2}{16\pi r^2}R^-_{\ell, \ell+\frac{1}{2},k}(r) &= k^2 R^+_{\ell+1, \ell+\frac{1}{2},k}(r) \ .
\end{split}
\end{equation}
The interacting potential $\mathcal{V}(\vec{r})$ not only connects $R^+$ and $R^-$ but also states with $\ell$ and $\ell\pm1$.

Next, we address the boundary conditions for the scattering problem. In the absence of any interactions $\Psi_k^\pm(\vec{r})$ are simply plane waves,
\begin{equation}
\Psi^\pm_k(\vec{r}) = e^{ikz} {\textstyle\left| \frac{1}{2}, \frac{1}{2}\right\rangle} = \sum_{\ell=0}^\infty i^\ell \sqrt{4\pi (2\ell+1)} j_\ell(kr) Y_{\ell,0}(\theta, \phi){\textstyle\left| \frac{1}{2}, \frac{1}{2}\right\rangle} \ .
\end{equation}
The fermion spin state is chosen such that the total angular momentum in the $\hat z$ direction is $m_j=\frac{1}{2}$, consistent with what was considered above.
Using the relations in Eq.~\eqref{eq:CB} and the asymptotic forms of the spherical Bessel function $j_\ell$,
the free wavefunctions $\Psi^\pm_k$ at $r\to\infty$ can be written as
\begin{equation}
\Psi^\pm_k(\vec{r}) \xrightarrow[]{r\to\infty} \sum_{l=0}^\infty \frac{\sqrt{4\pi (\ell+1)}}{kr} \left[ i^\ell \textstyle{\sin\left(kr-\frac{\pi\ell}{2}\right) \langle\theta,\phi\left|\ell, \frac{1}{2}, \ell+\frac{1}{2}, \frac{1}{2}\right\rangle} + i^{\ell+1} \sin\left(kr-\frac{\pi(\ell+1)}{2}\right) \textstyle{\langle\theta,\phi\left|\ell+1, \frac{1}{2}, \ell+\frac{1}{2}, \frac{1}{2}\right\rangle} \rule{0mm}{5mm}\right] \ , 
\end{equation}

Turning on an interacting potential in the scattering problem modifies the phase shift of each outgoing partial wave while keeping unitarity preserved. The interacting wavefunctions $\Psi^\pm_k$ must take the asymptotic forms
\begin{equation}\label{eq:phaseshifts}
\begin{split}
\Psi^\pm_k(\vec{r}) 
&\xrightarrow[]{r\to\infty}\sum_{l=0}^\infty \frac{\sqrt{4\pi (\ell+1)}}{kr} \left[ i^\ell e^{i\delta^\pm_{\ell,\ell+1/2}} \textstyle{\sin\left(kr-\frac{\pi\ell}{2}+ \delta^\pm_{\ell,\ell+1/2}\right) \langle\theta,\phi\left|\ell, \frac{1}{2}, \ell+\frac{1}{2}, \frac{1}{2}\right\rangle} \rule{0mm}{5mm}\right.\\
&\hspace{4cm}\left.+ i^{\ell+1} e^{i\delta^\pm_{\ell+1,\ell+1/2}} \textstyle{\sin\left(kr-\frac{\pi(\ell+1)}{2}+ \delta^\pm_{\ell+1,\ell+1/2}\right) \langle\theta,\phi\left|\ell+1, \frac{1}{2}, \ell+\frac{1}{2}, \frac{1}{2}\right\rangle} \rule{0mm}{5mm}\right]  \ , 
\end{split}
\end{equation}
where $\delta^\pm_{\ell,j}\ (j=\ell\pm1/2)$ are the phase shifts introduced for different $\ell$ and $j$ quantum numbers. In general, $\delta^+_{\ell,j}\neq \delta^-_{\ell,j}$.

The scattering cross sections can be expressed in terms of the phase shifts. We find
\begin{eqnarray}
\sigma &=& \frac{4\pi}{k^2} \sum_{\ell=0}^\infty (\ell+1) \left[ \sin^2 \delta^+_{\ell,\ell+1/2} + \sin^2 \delta^+_{\ell+1,\ell+1/2} \rule{0mm}{5mm}\right] \ , \label{eq:sigma}\\
\sigma_T &=& \frac{4\pi}{k^2} \sum_{\ell=0}^\infty \frac{\ell+1}{2\ell+3} \left[ (\ell+2)\sin^2 \left(\delta^+_{\ell,\ell+1/2}-\delta^+_{\ell+1,\ell+3/2}\right) + (\ell+2)\sin^2 \left( \delta^+_{\ell+1,\ell+1/2} - \delta^+_{\ell+2,\ell+3/2} \right) \rule{0mm}{5mm}\right.\nonumber\\
&&\hspace{2.5cm}+\left. \frac{1}{2\ell+1} \sin^2\left( \delta^+_{\ell,\ell+1/2} - \delta^+_{\ell+1,\ell+1/2} \right) \rule{0mm}{5mm}\right] \ , \label{eq:sigmaT}\\
\sigma_V &=& \frac{4\pi}{k^2} \sum_{\ell=0}^\infty \frac{\ell+1}{2\ell+3} \left\{ \frac{(\ell+2)(\ell+3)}{(2\ell+5)} \left[ \sin^2\left( \delta^+_{\ell,\ell+1/2} - \delta^+_{\ell+2, \ell+5/2} \right) + \sin^2\left( \delta^+_{\ell+1,\ell+1/2} - \delta^+_{\ell+3, \ell+5/2} \right) \rule{0mm}{4mm}\right] \rule{0mm}{5mm}\right. \nonumber \\
&&\left.\hspace{2.5cm}+\frac{2(\ell+2)}{(2\ell+1)(2\ell+5)} \left[ \sin^2\left( \delta^+_{\ell,\ell+1/2} - \delta^+_{\ell+2, \ell+3/2} \right) + \sin^2\left( \delta^+_{\ell+1, \ell+1/2} - \delta^+_{\ell+1,\ell+3/2}  \right)\right] \right\} \ .\label{eq:sigmaV}
\end{eqnarray}
These results are also valid if all $\delta_{\ell, j}^+$ are replaced by $\delta_{\ell, j}^-$. 
In the special case where the interacting potential and phase shift becomes $j$ independent, the above cross sections agrees with those given in~\cite{Colquhoun:2020adl}.

Before proceeding to solve the Schr\"odinger equations, a crucial observation is that dark matter mass in all the terms of the above Schr\"odinger equations have been canceled away and $k$ is the only remaining parameter with a dimension. 
Without dark matter mass, one cannot reduce $k$ to the dimensionless velocity.
As a result, the dimensionless phase shifts cannot depend on $k$ (or $v$) but rather the dimensionless parameters such as $y$ and $\ell$. This already leads us to conclude that all cross sections (Eqs.~\eqref{eq:sigma}, \eqref{eq:sigmaT}, \eqref{eq:sigmaV}) must have a common velocity dependence $\sim v^{-2}$ in the massless neutrino limit.

\section{Perturbation Theory}

The velocity independence of the scattering phase shifts suggests that the coupled Schr\"odinger equations \eqref{eq:SE1} and \eqref{eq:SE2} could be solved perturbatively. Due to the higher dimensional operator origin (see Eq.~\eqref{eq:Yukawa}), the Yukawa coupling $y$ must be less than $\mathcal{O}(1)$.
Furthermore, $y$ is bounded from above experimentally by various precision measurements of various Standard Model decay rates, as will be elaborated on next section. The perturbative expansion we resort to is based on the small parameter $y^2/(16\pi) \ll 1$.

Treating the potential term as a small perturbation, we expand the radial part of the scattering wavefunctions as
\begin{equation}
\begin{split}
R^\pm_{\ell, \ell+\frac{1}{2}, k}(r) &= R^{\pm(0)}_{\ell, \ell+\frac{1}{2}, k}(r) + R^{\pm(1)}_{\ell, \ell+\frac{1}{2}, k}(r) \ , \\
R^\pm_{\ell+1, \ell+\frac{1}{2}, k}(r) &= R^{\pm(0)}_{\ell+1, \ell+\frac{1}{2}, k}(r) + R^{\pm(1)}_{\ell+1, \ell+\frac{1}{2}, k}(r) \ ,
\end{split}
\end{equation}
where the zeroth order terms
\begin{equation}\label{eq:R0}
\begin{split}
R^{\pm(0)}_{\ell, \ell+\frac{1}{2}, k}(r) &= i^\ell \sqrt{4\pi (\ell+1)} j_\ell (kr) \ , \\
R^{\pm(0)}_{\ell+1, \ell+\frac{1}{2}, k}(r) &= i^{\ell+1} \sqrt{4\pi (\ell+1)} j_{\ell+1} (kr)\ ,
\end{split}
\end{equation}
correspond to the plane wave solutions,
and the first order terms $R^{\pm(1)}$ vanish in the $y\to0$ limit.

Interestingly, all $R^{\pm(1)}$ can be solved analytically by plugging the above $R^{\pm(0)}$ solutions into the potential term of \eqref{eq:SE1} and \eqref{eq:SE2}. This corresponds to the Born approximation~\cite{Born:1926yhp}. For the $R^+$ wavefunction, 
\begin{equation}\label{eq:R1}
\begin{split}
R^{+(1)}_{\ell, \ell+\frac{1}{2}, k}(r) &= - \frac{i^\ell y^2}{16\sqrt\pi \sqrt{\ell+1}} j_{\ell+1} (kr) \ , \\
R^{+(1)}_{\ell+1, \ell+\frac{1}{2}, k}(r) &= -\frac{i^{\ell+1}y^2}{16\sqrt\pi \sqrt{\ell+1}} j_{\ell} (kr)\ .
\end{split}
\end{equation}
The corresponding solutions for $R^-$ are the opposite.

Adding Eqs.~\eqref{eq:R0} and \eqref{eq:R1} and take the large $r$ limit, we get
\begin{equation}
\begin{split}
R^{\pm}_{\ell, \ell+\frac{1}{2}, k}(r)  &\simeq 
\frac{i^\ell \sqrt{4\pi(\ell+1)}}{kr} \left[ \sin\left(kr-\frac{\pi\ell}{2}\right) \pm \frac{y^2}{32\pi (\ell+1)} \cos\left(kr-\frac{\pi\ell}{2}\right) \right] \ , \\
R^{\pm}_{\ell+1, \ell+\frac{1}{2}, k}(r) &\simeq \frac{i^{\ell+1} \sqrt{4\pi(\ell+1)}}{kr} \left[ \sin\left(kr-\frac{\pi(\ell+1)}{2}\right) \mp \frac{y^2}{32\pi (\ell+1)} \cos\left(kr-\frac{\pi(\ell+1)}{2}\right) \right] 
\ .
\end{split}
\end{equation}
Matching these results to the asymptotic wavefunctions in Eq.~\eqref{eq:phaseshifts}, we obtain
\begin{equation}\label{eq:phases}
\delta_{l,l+\frac{1}{2}}^+ \simeq - \delta_{l+1,l+\frac{1}{2}}^+ \simeq - \delta_{l,l+\frac{1}{2}}^- \simeq \delta_{l+1,l+\frac{1}{2}}^- \simeq \frac{y^2}{32\pi(\ell+1)} \ ,
\end{equation}
which holds well for $y^2/(16\pi)\ll1$. The matching was not perfect but differences only occur at higher orders. This observation enables Eq.~\eqref{eq:phases} to hold as the leading result in the small $y^2/(16\pi)$ expansion.

Plugging the phase shift results back to Eqs.~\eqref{eq:sigma}, \eqref{eq:sigmaT}, \eqref{eq:sigmaV}, we find
\begin{equation}
\begin{split}
\sigma_T &\simeq \frac{4\pi}{k^2} \left(\frac{y^2}{32\pi}\right)^2 \sum_{l=0}^\infty \frac{2(4\ell+5)}{(\ell+1)(\ell+2)(2\ell+1)(2\ell+3)}  = \frac{y^4}{32\pi m^2 v^2} \ , \\
\sigma_V &\simeq \frac{4\pi}{k^2} \left(\frac{y^2}{32\pi}\right)^2 \sum_{l=0}^\infty \frac{4(4\ell+7)}{(\ell+2)(\ell+3)(2\ell+1)(2\ell+3)}  = \frac{y^4}{32\pi m^2 v^2} \ .
\end{split}
\end{equation}
In contrast, the total cross section
\begin{equation}
\sigma \simeq \frac{8\pi}{k^2} \left(\frac{y^2}{32\pi}\right)^2 \sum_{l=0}^\infty \frac{1}{\ell+1}\ , 
\end{equation}
is divergent. 

All these results are consistent with the findings based on tree-level Feynman diagram calculation, Eqs.~\eqref{eq:xsec0} and \eqref{eq:xsec} in the $m_\nu\to0$ limit. The perturbative solution of the Schr\"odinger equation allows us to conclude that unlike the case of a light vector or scalar boson mediated dark force, the Born approximation is sufficient and contributions from higher ladder diagrams are unimportant for light neutrino (fermion) mediated dark matter self-interaction.

\section{Dark Matter Phenomenology}

The analysis in the previous sections allows us to trust the tree-level result for the momentum-transfer and viscosity cross sections for $\chi\phi\to\chi\phi$, the key process for $\chi$ and $\phi$ (degenerate in mass) to play the role of a self-interacting dark matter. It is also useful to recall the associated assumption that the dark matter relic abundances are asymmetric, i.e., only the $\chi$ and $\phi$ particles are around in the universe today. Their antiparticles $\bar\chi, \phi^*$ have no population. In this section, we will derive the parameter space of the model where self-interacting dark matter can be realized and discuss the evolution of their number densities in the early universe. 

\begin{figure}[t]
  \begin{center}
  \includegraphics[width=0.618\textwidth]{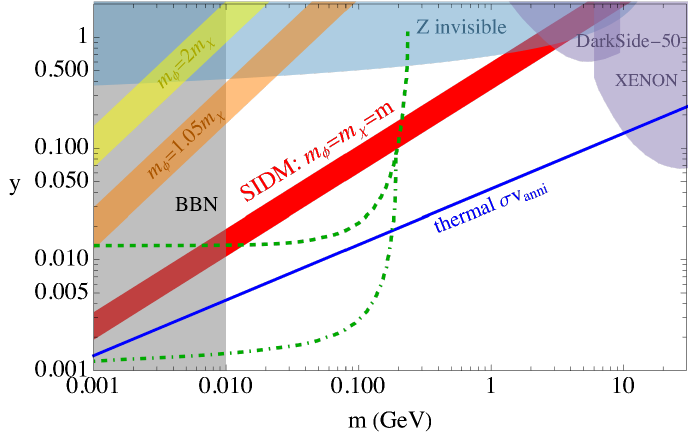}
  \end{center}
  \caption{Parameter space in the $y$ versus $m$ plane for self-interacting dark matter with $1 \,{\rm cm^2/g}< \sigma_T v/m < 10 \,{\rm cm^2/g}$. The red band is for the single-neutrino-exchange mediated interacting potential considered in this work where the dark scalar $\phi$ and fermion $\chi$ are degenerate in mass ($m_\chi=m_\phi=m$) and both serve as the dark matter.
  The yellow and orange bands are for non-degenerate case considered in~\cite{Orlofsky:2021mmy} where the interacting potential is generated via double-neutrino exchange. The blue line corresponds to the annihilation cross section needed for WIMP-like thermal dark matter. Asymmetric dark matter lives above the blue line.
  The gray, blue and purple shaded regions are excluded by BBN, invisible $Z$ decay width and dark matter direct detection, respectively, as detailed in the text. If the neutrino that couples to dark matter is of the $\nu_e$ ($\nu_\mu$) flavor, there are additional constraints from kaon decays that exclude the region above the dashed (dot-dashed) curve. 
  }
  \label{FIG2}
\end{figure}

\begin{figure}[t]
  \begin{center}  \includegraphics[width=0.618\textwidth]{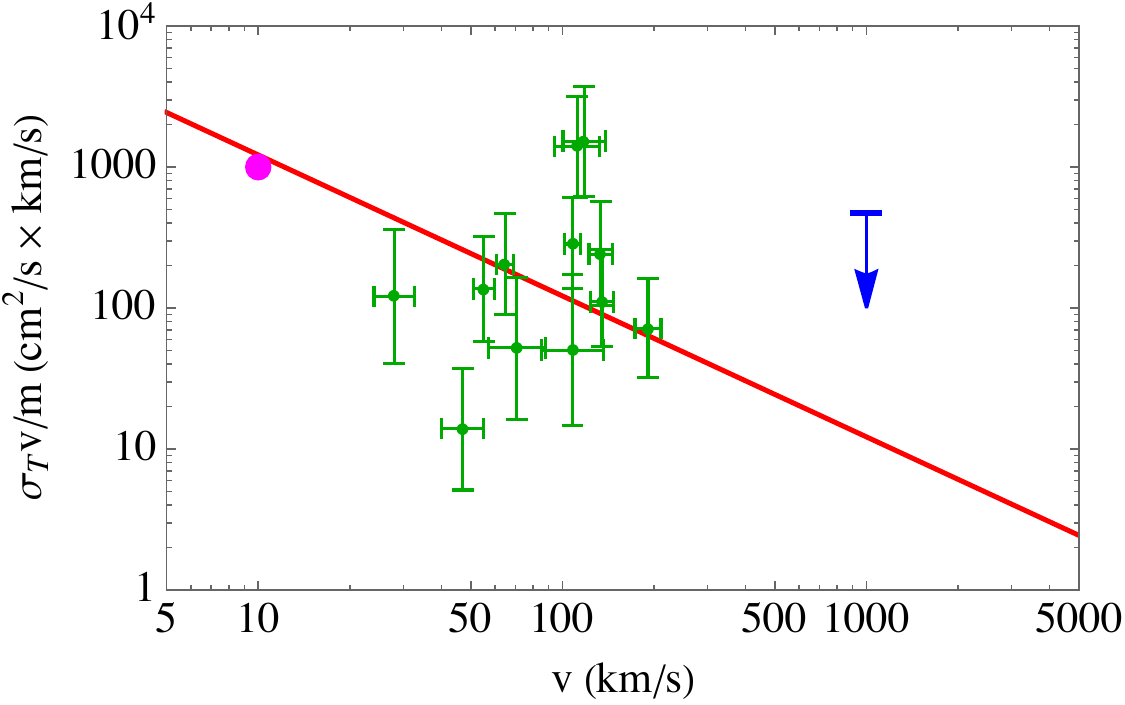}
  \end{center}
  \caption{The red curve shows the velocity dependence of the dark matter self-interaction cross section in the single-neutrino-exchange dark force model considered in this work, with $y=0.5$ and $m=1$\,GeV. The green points with error bars indicate the favored parameter space for addressing the core-or-cusp problem in dwarf galaxies and low surface brightness spiral galaxies~\cite{Kaplinghat:2015aga}. The blue bar with arrow indicates the upper bound on dark matter self-interaction cross section set by colliding galaxy cluster observations~\cite{Harvey:2015hha}. The Magenta disk corresponds to a favored point used in a recent analysis of Milky Way dwarf galaxies. 
  }
  \label{FIG3}
\end{figure}

In each panel of Fig.~\ref{FIG2}, the red band shows 
the self-interacting dark matter favored parameter space in the $y$ versus $m$ plane, with
\begin{equation}
\frac{\sigma_T}{m} = 1 -10 \,{\rm cm^2/g} \ ,
\end{equation}
where $\sigma_T$ is given in Eq.~\eqref{eq:xsecTV} and we use
a typical dark matter velocity in dwarf galaxies $v\sim 50\,{\rm km/s}$ (c.f. Fig.~\ref{FIG3}). For dark matter mass above MeV scale, its typical three-momentum is much higher than the neutrino mass scale. To very good approximation, we are working in the massless neutrino limit in this figure.

For comparison, the yellow and orange bands show the self-interacting dark matter parameter space for non-degenerate $\chi$ and $\phi$, with $m_\phi=2m$ and $m_\phi=1.05 m$, respectively. In both cases, only $\chi$ is the dark matter (with mass $m$). The $\phi$-$\chi$ mass difference is much higher than dark matter's kinetic energy in galaxies. As a result, $\phi$ can be integrated out in the dark matter scattering problem. This scenario has been explored in detail in~\cite{Orlofsky:2021mmy}, where dark matter self-interaction must occur via double-neutrino exchange at loop level. This leads to a $1/r^5$ repulsive potential among the $\chi$ particles. The corresponding $\chi\chi\to\chi\chi$ scattering cross section is
\begin{equation}
\sigma_T \simeq \sigma \simeq \frac{y^{8/3} m^{2/3}}{24\sqrt[3]{6} [\Gamma\left(\frac{7}{6}\right)]^2 (m_\phi^2-m^2)^{4/3}} \ ,
\end{equation}
where $\Gamma$ is the Euler gamma function.
Without the $1/v^2$ enhancement factor in the cross section, a much larger coupling $y$ is needed to realize the same dark matter self-interaction cross section. 
By reducing the $\chi$-$\phi$ mass difference, the bands move downwards, until $\phi$ is sufficient light and liberated from the loop. The latter case corresponds to the single-neutrino exchange mediated dark force considered in the present work.
The self-interacting dark matter parameter space eventually saturates to the red band.

Back to the degenerate case, let us consider a thermal history of the universe.
For $\chi, \phi$ to be the asymmetric dark matter, they must be able to efficiently annihilate away the antiparticles when the temperature of the universe falls well below their masses~\cite{Kaplan:2009ag}. The annihilation cross sections from their Yukawa coupling with neutrino (Eq.~\eqref{eq:Yukawa}) are
\begin{equation}
(\sigma v_{\rm Møl})_{\chi\bar\chi\to\nu\bar\nu} = \frac{y^4}{128\pi m^2} \ , \quad (\sigma v_{\rm Møl})_{\phi\phi^*\to\nu\bar\nu} = \frac{y^4 v_{\rm Møl}^2}{192\pi m^2} \ ,
\end{equation}
where $v_{\rm Møl}$ is the dark matter velocity during the freeze out. The second process is slower and $P$-wave suppressed. 
The blue line in Fig.~\ref{FIG2} corresponds to a thermal averaged cross section $\langle\sigma v_{\rm Møl}\rangle_{\phi\phi^*\to\nu\bar\nu}$ that can produce the observed relic density for WIMP-like dark matter (no asymmetry) via the freeze out mechanism~\cite{Kolb:1990vq}, where $\langle v_{\rm Møl}^2\rangle \simeq 6T/m$. 
The parameter space consistent with asymmetric dark matter is the region well above the blue line. 
We attribute the origin of the dark asymmetries to high-scale physics beyond the simple model considered here. 

The gray shaded region where $m\lesssim10$\,MeV is excluded due to the excessive contribution from $\chi, \phi$ to $\Delta N_{\rm eff}$ during BBN. Clearly, in the entire BBN-allowed region, the self-interacting dark matter (red bad) always has sufficiently large annihilation cross sections for dark matter to be asymmetric.

For dark matter mass above a few GeV, direct detection constraints become relevant. In this model, dark matter $\chi$ can elastically scatter off a nucleus target via a $\nu, \phi$ loop and the $Z$-boson exchange. The spin-independent $\chi$-neutron scattering cross section is~\cite{Batell:2017cmf}
\begin{equation}
\sigma_{\chi n} \simeq \frac{y^4 G_F^2 \mu_{\chi n}^2}{2048\pi^5}\left(\log\frac{\Lambda}{m}\right)^2 \ ,
\end{equation}
where $\mu_{\chi n}$ is the reduced mass and we approximated $\sin^2\theta_W\simeq1/4$. The loop diagram is logarithmic divergent and we regularize it with the same cutoff scale $\Lambda$ in Eq.~\eqref{eq:LAM} where $\Lambda^2 = (2\sqrt{2} G_F y^2)^{-1}$.
The most recent results from the XENONnT~\cite{XENON:2023cxc} and DarkSide-50~\cite{DarkSide-50:2022qzh} experiments exclude the purple region in the upper-right corner. 
They set an upper bound on the self-interacting dark matter mass of $\sim3\,$GeV in this mdoel.
Upcoming low-mass dark matter (GeV scale and below) direct detection experiments could be used to constrain the self-interacting dark matter parameter space (red band).

The Yukawa coupling $y$ is bounded from above by the invisible $Z$-boson decay width. In the presence of light $\chi$ and $\phi$, the $Z$ boson can decay into $\chi \phi^*\bar \nu$ and $\bar\chi \phi \nu$ final states. The corresponding partial decay rate has been calculated in~\cite{Orlofsky:2021mmy}. In the $M_Z\gg m$ limit, 
\begin{equation}
\Gamma_{Z\to \chi\phi^*\bar\nu} = \Gamma_{Z\to \bar \chi\phi\nu} \simeq \frac{y^2 G_F M_Z^3}{192\sqrt{2}\pi^3} \left[\log\left(\frac{M_Z}{m}\right) - \frac{5}{3}\right] \ .
\end{equation}
Using the total $Z$ width $2.5\,$GeV and the invisible branching ratio ${\rm Br}(Z\to{\rm invisible})=(20\pm0.055)\%$, the $2\sigma$ exclusion corresponds to the blue shaded region in Fig.~\ref{FIG2}.

All the above results are independent of the neutrino flavor $\chi$ and $\phi$ couple to. If the neutrino is $\nu_e$ or $\nu_\mu$, there are additional flavor-dependent constraints from charged meson $\mathfrak{m}=(\pi, K)$ leptonic decays where the final state neutrino can turn into $\chi+\phi^*$ and remain invisible. The corresponding decay rate is
\begin{equation}
\Gamma_{\mathfrak{m}^+\to \ell^+\chi\phi^*} = \frac{y^2 G_F^2 m_\mathfrak{m}^3 F_\mathfrak{m}^2}{256\pi^3} \int_{4z}^{(1-\sqrt{y})^2} dx \frac{\sqrt{x-4z}\sqrt{x^2 -2x(y+1)+(y-1)^2}(x+y-(x-y)^2)}{x^{3/2}} \ ,
\end{equation}
where $y=m_\ell^2/m_\mathfrak{m}^2$, $z=m^2/m_\mathfrak{m}^2$, and $F_\mathfrak{m}$ is the meson decay constant. We reinterpret the upper limits on the branching ratios
${\rm Br}(K^+\to e^+ 3\nu)<6\times10^{-5}$~\cite{Heintze:1977kk} and ${\rm Br}(K^+\to \mu^+ 3\nu)<1.0\times10^{-6}$~\cite{NA62:2021bji} to constrain the above decay modes into dark matter. The resulting upper bound on $y$ are shown by the green dashed and dot-dashed curves, for $\nu_e$ and $\nu_\mu$ cases, respectively. 
Constraints from similar charged pion decay channels~\cite{PIENU:2020las} are less important.
In these cases, the self-interacting dark matter mass is constrained to be above 200 MeV.

In Fig.~\ref{FIG3}, the red curve presents the velocity dependence of the dark matter self-interaction cross sections. We choose a dark matter mass $m=0.5$\,GeV and dark-matter-neutrino Yukawa coupling $y=0.35$. (The curve stays roughly the same for other points along the red band in Fig.~\ref{FIG2}.) Because the largest neutrino mass is constrained by cosmology to be $\lesssim 0.1\,$eV~\cite{Planck:2018vyg}, over a wide range of dark matter velocities relevant for small and large scales, $\sigma_T v$ and $\sigma_V v$ both scale as $1/v$ and are shown by the red curve. 
Meanwhile, they are always safely below the geometric cross section set by the de Briglie wavelength $\sim (mv)^{-2}$, as long as the Yukawa coupling $y$ is perturbative.

Such a velocity dependence of the cross sections can accommodate a self-interacting dark matter that works well for both small and large scale structures of the universe.
In Fig.~\ref{FIG3}, the green points, taken from~\cite{Kaplinghat:2015aga}, indicate the favored cross section values that can solve the core-or-cusp problem for dwarf galaxies and low surface brightness spiral galaxies. The corresponding dark matter velocity lie in the range of $20-200$\,km/s. 
On the other hand, the self-interaction cross section is also bounded from above by the observation of colliding galaxy clusters~\cite{Harvey:2015hha}, $\sigma_T/m < 0.47\,{\rm cm^2/g}$, or $\sigma_T v/m < 470\,{\rm cm^2/g \times km/s}$. In this case, the dark matter velocities are higher, $v\sim 1000\,$km/s and the upper bound is shown by the blue arrow.

A recent simulation of self-interacting dark matter takes into account of environmental effects of dwarf galaxies orbiting the Milky Way and suggests that a self-interaction
cross section of order $\sigma_T/m=100\,{\rm cm^2/g}$ at velocity $10\,$km/s is favored~\cite{Yang:2022mxl}. It corresponds to $\sigma_T v/m=1000\,{\rm cm^2/g\times km/s}$, as shown by the magenta point in Fig.~\ref{FIG3}. Our single-neutrino mediated dark force model can also fit this point well.

Due to the $1/v^2$ dependence, the neutrino-mediated dark matter self-interaction cross section continues to grow at even lower velocities, until $v$ falls below $m_\nu/m \sim 10^{-9}c (m_\nu/0.1\,{\rm eV})(100\,{\rm MeV}/m)$.
This feature allows it to be distinguished from other models where the self-interaction is mediated by a massive/massless dark photon/scalar or via a Breit-Wigner resonance.
Moreover, if there exists a window with $v \gtrsim\sqrt{m_\nu/m}$ and the $2\to3$ body scattering processes $\chi\phi\to\phi\phi\nu$ and $\chi\phi\to\chi\chi\bar\nu$ are sufficiently fast, dark matter kinetic energy can be dissipated away via the radiation of neutrinos.
It would be interesting to explore the observable consequences of these strong self-interactions of dark matter in low velocity astrophysical environments.

\section{Acknowledgements}
I thank Flip Tanedo, Walter Tangarife, Hai-Bo Yu, and Jiapeng Zhang for useful discussions and communications. 
This work is supported by a Subatomic Physics Discovery Grant (individual) from the Natural Sciences and Engineering Research Council of Canada, and by the Arthur B. McDonald Canadian Astroparticle Physics Research Institute.

\bibliography{NuSIDM.References}

\begin{thebibliography}{52}%
\makeatletter
\providecommand \@ifxundefined [1]{%
 \@ifx{#1\undefined}
}%
\providecommand \@ifnum [1]{%
 \ifnum #1\expandafter \@firstoftwo
 \else \expandafter \@secondoftwo
 \fi
}%
\providecommand \@ifx [1]{%
 \ifx #1\expandafter \@firstoftwo
 \else \expandafter \@secondoftwo
 \fi
}%
\providecommand \natexlab [1]{#1}%
\providecommand \enquote  [1]{``#1''}%
\providecommand \bibnamefont  [1]{#1}%
\providecommand \bibfnamefont [1]{#1}%
\providecommand \citenamefont [1]{#1}%
\providecommand \href@noop [0]{\@secondoftwo}%
\providecommand \href [0]{\begingroup \@sanitize@url \@href}%
\providecommand \@href[1]{\@@startlink{#1}\@@href}%
\providecommand \@@href[1]{\endgroup#1\@@endlink}%
\providecommand \@sanitize@url [0]{\catcode `\\12\catcode `\$12\catcode
  `\&12\catcode `\#12\catcode `\^12\catcode `\_12\catcode `\%12\relax}%
\providecommand \@@startlink[1]{}%
\providecommand \@@endlink[0]{}%
\providecommand \url  [0]{\begingroup\@sanitize@url \@url }%
\providecommand \@url [1]{\endgroup\@href {#1}{\urlprefix }}%
\providecommand \urlprefix  [0]{URL }%
\providecommand \Eprint [0]{\href }%
\providecommand \doibase [0]{http://dx.doi.org/}%
\providecommand \selectlanguage [0]{\@gobble}%
\providecommand \bibinfo  [0]{\@secondoftwo}%
\providecommand \bibfield  [0]{\@secondoftwo}%
\providecommand \translation [1]{[#1]}%
\providecommand \BibitemOpen [0]{}%
\providecommand \bibitemStop [0]{}%
\providecommand \bibitemNoStop [0]{.\EOS\space}%
\providecommand \EOS [0]{\spacefactor3000\relax}%
\providecommand \BibitemShut  [1]{\csname bibitem#1\endcsname}%
\let\auto@bib@innerbib\@empty
\bibitem [{\citenamefont {Spergel}\ and\ \citenamefont
  {Steinhardt}(2000)}]{Spergel:1999mh}%
  \BibitemOpen
  \bibfield  {author} {\bibinfo {author} {\bibfnamefont {David~N.}\
  \bibnamefont {Spergel}}\ and\ \bibinfo {author} {\bibfnamefont {Paul~J.}\
  \bibnamefont {Steinhardt}},\ }\bibfield  {title} {\enquote {\bibinfo {title}
  {{Observational evidence for selfinteracting cold dark matter}},}\ }\href
  {\doibase 10.1103/PhysRevLett.84.3760} {\bibfield  {journal} {\bibinfo
  {journal} {Phys. Rev. Lett.}\ }\textbf {\bibinfo {volume} {84}},\ \bibinfo
  {pages} {3760--3763} (\bibinfo {year} {2000})},\ \Eprint
  {http://arxiv.org/abs/astro-ph/9909386} {arXiv:astro-ph/9909386} \BibitemShut
  {NoStop}%
\bibitem [{\citenamefont {Dave}\ \emph {et~al.}(2001)\citenamefont {Dave},
  \citenamefont {Spergel}, \citenamefont {Steinhardt},\ and\ \citenamefont
  {Wandelt}}]{Dave:2000ar}%
  \BibitemOpen
  \bibfield  {author} {\bibinfo {author} {\bibfnamefont {Romeel}\ \bibnamefont
  {Dave}}, \bibinfo {author} {\bibfnamefont {David~N.}\ \bibnamefont
  {Spergel}}, \bibinfo {author} {\bibfnamefont {Paul~J.}\ \bibnamefont
  {Steinhardt}}, \ and\ \bibinfo {author} {\bibfnamefont {Benjamin~D.}\
  \bibnamefont {Wandelt}},\ }\bibfield  {title} {\enquote {\bibinfo {title}
  {{Halo properties in cosmological simulations of selfinteracting cold dark
  matter}},}\ }\href {\doibase 10.1086/318417} {\bibfield  {journal} {\bibinfo
  {journal} {Astrophys. J.}\ }\textbf {\bibinfo {volume} {547}},\ \bibinfo
  {pages} {574--589} (\bibinfo {year} {2001})},\ \Eprint
  {http://arxiv.org/abs/astro-ph/0006218} {arXiv:astro-ph/0006218} \BibitemShut
  {NoStop}%
\bibitem [{\citenamefont {Tulin}\ and\ \citenamefont
  {Yu}(2018)}]{Tulin:2017ara}%
  \BibitemOpen
  \bibfield  {author} {\bibinfo {author} {\bibfnamefont {Sean}\ \bibnamefont
  {Tulin}}\ and\ \bibinfo {author} {\bibfnamefont {Hai-Bo}\ \bibnamefont
  {Yu}},\ }\bibfield  {title} {\enquote {\bibinfo {title} {{Dark Matter
  Self-interactions and Small Scale Structure}},}\ }\href {\doibase
  10.1016/j.physrep.2017.11.004} {\bibfield  {journal} {\bibinfo  {journal}
  {Phys. Rept.}\ }\textbf {\bibinfo {volume} {730}},\ \bibinfo {pages} {1--57}
  (\bibinfo {year} {2018})},\ \Eprint {http://arxiv.org/abs/1705.02358}
  {arXiv:1705.02358 [hep-ph]} \BibitemShut {NoStop}%
\bibitem [{\citenamefont {Adhikari}\ \emph {et~al.}(2022)\citenamefont
  {Adhikari} \emph {et~al.}}]{Adhikari:2022sbh}%
  \BibitemOpen
  \bibfield  {author} {\bibinfo {author} {\bibfnamefont {Susmita}\ \bibnamefont
  {Adhikari}} \emph {et~al.},\ }\bibfield  {title} {\enquote {\bibinfo {title}
  {{Astrophysical Tests of Dark Matter Self-Interactions}},}\ }\href@noop {} {\
   (\bibinfo {year} {2022})},\ \Eprint {http://arxiv.org/abs/2207.10638}
  {arXiv:2207.10638 [astro-ph.CO]} \BibitemShut {NoStop}%
\bibitem [{\citenamefont {Feng}\ \emph {et~al.}(2009)\citenamefont {Feng},
  \citenamefont {Kaplinghat}, \citenamefont {Tu},\ and\ \citenamefont
  {Yu}}]{Feng:2009mn}%
  \BibitemOpen
  \bibfield  {author} {\bibinfo {author} {\bibfnamefont {Jonathan~L.}\
  \bibnamefont {Feng}}, \bibinfo {author} {\bibfnamefont {Manoj}\ \bibnamefont
  {Kaplinghat}}, \bibinfo {author} {\bibfnamefont {Huitzu}\ \bibnamefont {Tu}},
  \ and\ \bibinfo {author} {\bibfnamefont {Hai-Bo}\ \bibnamefont {Yu}},\
  }\bibfield  {title} {\enquote {\bibinfo {title} {{Hidden Charged Dark
  Matter}},}\ }\href {\doibase 10.1088/1475-7516/2009/07/004} {\bibfield
  {journal} {\bibinfo  {journal} {JCAP}\ }\textbf {\bibinfo {volume} {07}},\
  \bibinfo {pages} {004} (\bibinfo {year} {2009})},\ \Eprint
  {http://arxiv.org/abs/0905.3039} {arXiv:0905.3039 [hep-ph]} \BibitemShut
  {NoStop}%
\bibitem [{\citenamefont {Buckley}\ and\ \citenamefont
  {Fox}(2010)}]{Buckley:2009in}%
  \BibitemOpen
  \bibfield  {author} {\bibinfo {author} {\bibfnamefont {Matthew~R.}\
  \bibnamefont {Buckley}}\ and\ \bibinfo {author} {\bibfnamefont {Patrick~J.}\
  \bibnamefont {Fox}},\ }\bibfield  {title} {\enquote {\bibinfo {title} {{Dark
  Matter Self-Interactions and Light Force Carriers}},}\ }\href {\doibase
  10.1103/PhysRevD.81.083522} {\bibfield  {journal} {\bibinfo  {journal} {Phys.
  Rev. D}\ }\textbf {\bibinfo {volume} {81}},\ \bibinfo {pages} {083522}
  (\bibinfo {year} {2010})},\ \Eprint {http://arxiv.org/abs/0911.3898}
  {arXiv:0911.3898 [hep-ph]} \BibitemShut {NoStop}%
\bibitem [{\citenamefont {Loeb}\ and\ \citenamefont
  {Weiner}(2011)}]{Loeb:2010gj}%
  \BibitemOpen
  \bibfield  {author} {\bibinfo {author} {\bibfnamefont {Abraham}\ \bibnamefont
  {Loeb}}\ and\ \bibinfo {author} {\bibfnamefont {Neal}\ \bibnamefont
  {Weiner}},\ }\bibfield  {title} {\enquote {\bibinfo {title} {{Cores in Dwarf
  Galaxies from Dark Matter with a Yukawa Potential}},}\ }\href {\doibase
  10.1103/PhysRevLett.106.171302} {\bibfield  {journal} {\bibinfo  {journal}
  {Phys. Rev. Lett.}\ }\textbf {\bibinfo {volume} {106}},\ \bibinfo {pages}
  {171302} (\bibinfo {year} {2011})},\ \Eprint {http://arxiv.org/abs/1011.6374}
  {arXiv:1011.6374 [astro-ph.CO]} \BibitemShut {NoStop}%
\bibitem [{\citenamefont {van~den Aarssen}\ \emph {et~al.}(2012)\citenamefont
  {van~den Aarssen}, \citenamefont {Bringmann},\ and\ \citenamefont
  {Pfrommer}}]{Aarssen:2012fx}%
  \BibitemOpen
  \bibfield  {author} {\bibinfo {author} {\bibfnamefont {Laura~G.}\
  \bibnamefont {van~den Aarssen}}, \bibinfo {author} {\bibfnamefont {Torsten}\
  \bibnamefont {Bringmann}}, \ and\ \bibinfo {author} {\bibfnamefont
  {Christoph}\ \bibnamefont {Pfrommer}},\ }\bibfield  {title} {\enquote
  {\bibinfo {title} {{Is dark matter with long-range interactions a solution to
  all small-scale problems of \textbackslash{}Lambda CDM cosmology?}}}\ }\href
  {\doibase 10.1103/PhysRevLett.109.231301} {\bibfield  {journal} {\bibinfo
  {journal} {Phys. Rev. Lett.}\ }\textbf {\bibinfo {volume} {109}},\ \bibinfo
  {pages} {231301} (\bibinfo {year} {2012})},\ \Eprint
  {http://arxiv.org/abs/1205.5809} {arXiv:1205.5809 [astro-ph.CO]} \BibitemShut
  {NoStop}%
\bibitem [{\citenamefont {Tulin}\ \emph
  {et~al.}(2013{\natexlab{a}})\citenamefont {Tulin}, \citenamefont {Yu},\ and\
  \citenamefont {Zurek}}]{Tulin:2012wi}%
  \BibitemOpen
  \bibfield  {author} {\bibinfo {author} {\bibfnamefont {Sean}\ \bibnamefont
  {Tulin}}, \bibinfo {author} {\bibfnamefont {Hai-Bo}\ \bibnamefont {Yu}}, \
  and\ \bibinfo {author} {\bibfnamefont {Kathryn~M.}\ \bibnamefont {Zurek}},\
  }\bibfield  {title} {\enquote {\bibinfo {title} {{Resonant Dark Forces and
  Small Scale Structure}},}\ }\href {\doibase 10.1103/PhysRevLett.110.111301}
  {\bibfield  {journal} {\bibinfo  {journal} {Phys. Rev. Lett.}\ }\textbf
  {\bibinfo {volume} {110}},\ \bibinfo {pages} {111301} (\bibinfo {year}
  {2013}{\natexlab{a}})},\ \Eprint {http://arxiv.org/abs/1210.0900}
  {arXiv:1210.0900 [hep-ph]} \BibitemShut {NoStop}%
\bibitem [{\citenamefont {Tulin}\ \emph
  {et~al.}(2013{\natexlab{b}})\citenamefont {Tulin}, \citenamefont {Yu},\ and\
  \citenamefont {Zurek}}]{Tulin:2013teo}%
  \BibitemOpen
  \bibfield  {author} {\bibinfo {author} {\bibfnamefont {Sean}\ \bibnamefont
  {Tulin}}, \bibinfo {author} {\bibfnamefont {Hai-Bo}\ \bibnamefont {Yu}}, \
  and\ \bibinfo {author} {\bibfnamefont {Kathryn~M.}\ \bibnamefont {Zurek}},\
  }\bibfield  {title} {\enquote {\bibinfo {title} {{Beyond Collisionless Dark
  Matter: Particle Physics Dynamics for Dark Matter Halo Structure}},}\ }\href
  {\doibase 10.1103/PhysRevD.87.115007} {\bibfield  {journal} {\bibinfo
  {journal} {Phys. Rev. D}\ }\textbf {\bibinfo {volume} {87}},\ \bibinfo
  {pages} {115007} (\bibinfo {year} {2013}{\natexlab{b}})},\ \Eprint
  {http://arxiv.org/abs/1302.3898} {arXiv:1302.3898 [hep-ph]} \BibitemShut
  {NoStop}%
\bibitem [{\citenamefont {Bellazzini}\ \emph {et~al.}(2013)\citenamefont
  {Bellazzini}, \citenamefont {Cliche},\ and\ \citenamefont
  {Tanedo}}]{Bellazzini:2013foa}%
  \BibitemOpen
  \bibfield  {author} {\bibinfo {author} {\bibfnamefont {Brando}\ \bibnamefont
  {Bellazzini}}, \bibinfo {author} {\bibfnamefont {Mathieu}\ \bibnamefont
  {Cliche}}, \ and\ \bibinfo {author} {\bibfnamefont {Philip}\ \bibnamefont
  {Tanedo}},\ }\bibfield  {title} {\enquote {\bibinfo {title} {{Effective
  theory of self-interacting dark matter}},}\ }\href {\doibase
  10.1103/PhysRevD.88.083506} {\bibfield  {journal} {\bibinfo  {journal} {Phys.
  Rev. D}\ }\textbf {\bibinfo {volume} {88}},\ \bibinfo {pages} {083506}
  (\bibinfo {year} {2013})},\ \Eprint {http://arxiv.org/abs/1307.1129}
  {arXiv:1307.1129 [hep-ph]} \BibitemShut {NoStop}%
\bibitem [{\citenamefont {Boddy}\ \emph {et~al.}(2014)\citenamefont {Boddy},
  \citenamefont {Feng}, \citenamefont {Kaplinghat},\ and\ \citenamefont
  {Tait}}]{Boddy:2014yra}%
  \BibitemOpen
  \bibfield  {author} {\bibinfo {author} {\bibfnamefont {Kimberly~K.}\
  \bibnamefont {Boddy}}, \bibinfo {author} {\bibfnamefont {Jonathan~L.}\
  \bibnamefont {Feng}}, \bibinfo {author} {\bibfnamefont {Manoj}\ \bibnamefont
  {Kaplinghat}}, \ and\ \bibinfo {author} {\bibfnamefont {Tim M.~P.}\
  \bibnamefont {Tait}},\ }\bibfield  {title} {\enquote {\bibinfo {title}
  {{Self-Interacting Dark Matter from a Non-Abelian Hidden Sector}},}\ }\href
  {\doibase 10.1103/PhysRevD.89.115017} {\bibfield  {journal} {\bibinfo
  {journal} {Phys. Rev. D}\ }\textbf {\bibinfo {volume} {89}},\ \bibinfo
  {pages} {115017} (\bibinfo {year} {2014})},\ \Eprint
  {http://arxiv.org/abs/1402.3629} {arXiv:1402.3629 [hep-ph]} \BibitemShut
  {NoStop}%
\bibitem [{\citenamefont {Hochberg}\ \emph {et~al.}(2015)\citenamefont
  {Hochberg}, \citenamefont {Kuflik}, \citenamefont {Murayama}, \citenamefont
  {Volansky},\ and\ \citenamefont {Wacker}}]{Hochberg:2014kqa}%
  \BibitemOpen
  \bibfield  {author} {\bibinfo {author} {\bibfnamefont {Yonit}\ \bibnamefont
  {Hochberg}}, \bibinfo {author} {\bibfnamefont {Eric}\ \bibnamefont {Kuflik}},
  \bibinfo {author} {\bibfnamefont {Hitoshi}\ \bibnamefont {Murayama}},
  \bibinfo {author} {\bibfnamefont {Tomer}\ \bibnamefont {Volansky}}, \ and\
  \bibinfo {author} {\bibfnamefont {Jay~G.}\ \bibnamefont {Wacker}},\
  }\bibfield  {title} {\enquote {\bibinfo {title} {{Model for Thermal Relic
  Dark Matter of Strongly Interacting Massive Particles}},}\ }\href {\doibase
  10.1103/PhysRevLett.115.021301} {\bibfield  {journal} {\bibinfo  {journal}
  {Phys. Rev. Lett.}\ }\textbf {\bibinfo {volume} {115}},\ \bibinfo {pages}
  {021301} (\bibinfo {year} {2015})},\ \Eprint {http://arxiv.org/abs/1411.3727}
  {arXiv:1411.3727 [hep-ph]} \BibitemShut {NoStop}%
\bibitem [{\citenamefont {Soni}\ and\ \citenamefont
  {Zhang}(2016)}]{Soni:2016gzf}%
  \BibitemOpen
  \bibfield  {author} {\bibinfo {author} {\bibfnamefont {Amarjit}\ \bibnamefont
  {Soni}}\ and\ \bibinfo {author} {\bibfnamefont {Yue}\ \bibnamefont {Zhang}},\
  }\bibfield  {title} {\enquote {\bibinfo {title} {{Hidden SU(N) Glueball Dark
  Matter}},}\ }\href {\doibase 10.1103/PhysRevD.93.115025} {\bibfield
  {journal} {\bibinfo  {journal} {Phys. Rev. D}\ }\textbf {\bibinfo {volume}
  {93}},\ \bibinfo {pages} {115025} (\bibinfo {year} {2016})},\ \Eprint
  {http://arxiv.org/abs/1602.00714} {arXiv:1602.00714 [hep-ph]} \BibitemShut
  {NoStop}%
\bibitem [{\citenamefont {Zhang}(2017)}]{Zhang:2016dck}%
  \BibitemOpen
  \bibfield  {author} {\bibinfo {author} {\bibfnamefont {Yue}\ \bibnamefont
  {Zhang}},\ }\bibfield  {title} {\enquote {\bibinfo {title} {{Self-interacting
  Dark Matter Without Direct Detection Constraints}},}\ }\href {\doibase
  10.1016/j.dark.2016.12.003} {\bibfield  {journal} {\bibinfo  {journal} {Phys.
  Dark Univ.}\ }\textbf {\bibinfo {volume} {15}},\ \bibinfo {pages} {82--89}
  (\bibinfo {year} {2017})},\ \Eprint {http://arxiv.org/abs/1611.03492}
  {arXiv:1611.03492 [hep-ph]} \BibitemShut {NoStop}%
\bibitem [{\citenamefont {Blennow}\ \emph {et~al.}(2017)\citenamefont
  {Blennow}, \citenamefont {Clementz},\ and\ \citenamefont
  {Herrero-Garcia}}]{Blennow:2016gde}%
  \BibitemOpen
  \bibfield  {author} {\bibinfo {author} {\bibfnamefont {Mattias}\ \bibnamefont
  {Blennow}}, \bibinfo {author} {\bibfnamefont {Stefan}\ \bibnamefont
  {Clementz}}, \ and\ \bibinfo {author} {\bibfnamefont {Juan}\ \bibnamefont
  {Herrero-Garcia}},\ }\bibfield  {title} {\enquote {\bibinfo {title}
  {{Self-interacting inelastic dark matter: A viable solution to the small
  scale structure problems}},}\ }\href {\doibase 10.1088/1475-7516/2017/03/048}
  {\bibfield  {journal} {\bibinfo  {journal} {JCAP}\ }\textbf {\bibinfo
  {volume} {03}},\ \bibinfo {pages} {048} (\bibinfo {year} {2017})},\ \Eprint
  {http://arxiv.org/abs/1612.06681} {arXiv:1612.06681 [hep-ph]} \BibitemShut
  {NoStop}%
\bibitem [{\citenamefont {Soni}\ \emph {et~al.}(2017)\citenamefont {Soni},
  \citenamefont {Xiao},\ and\ \citenamefont {Zhang}}]{Soni:2017nlm}%
  \BibitemOpen
  \bibfield  {author} {\bibinfo {author} {\bibfnamefont {Amarjit}\ \bibnamefont
  {Soni}}, \bibinfo {author} {\bibfnamefont {Huangyu}\ \bibnamefont {Xiao}}, \
  and\ \bibinfo {author} {\bibfnamefont {Yue}\ \bibnamefont {Zhang}},\
  }\bibfield  {title} {\enquote {\bibinfo {title} {{Cosmic selection rule for
  the glueball dark matter relic density}},}\ }\href {\doibase
  10.1103/PhysRevD.96.083514} {\bibfield  {journal} {\bibinfo  {journal} {Phys.
  Rev. D}\ }\textbf {\bibinfo {volume} {96}},\ \bibinfo {pages} {083514}
  (\bibinfo {year} {2017})},\ \Eprint {http://arxiv.org/abs/1704.02347}
  {arXiv:1704.02347 [hep-ph]} \BibitemShut {NoStop}%
\bibitem [{\citenamefont {McDermott}(2018)}]{McDermott:2017vyk}%
  \BibitemOpen
  \bibfield  {author} {\bibinfo {author} {\bibfnamefont {Samuel~D.}\
  \bibnamefont {McDermott}},\ }\bibfield  {title} {\enquote {\bibinfo {title}
  {{Is Self-Interacting Dark Matter Undergoing Dark Fusion?}}}\ }\href
  {\doibase 10.1103/PhysRevLett.120.221806} {\bibfield  {journal} {\bibinfo
  {journal} {Phys. Rev. Lett.}\ }\textbf {\bibinfo {volume} {120}},\ \bibinfo
  {pages} {221806} (\bibinfo {year} {2018})},\ \Eprint
  {http://arxiv.org/abs/1711.00857} {arXiv:1711.00857 [hep-ph]} \BibitemShut
  {NoStop}%
\bibitem [{\citenamefont {Chu}\ \emph {et~al.}(2019)\citenamefont {Chu},
  \citenamefont {Garcia-Cely},\ and\ \citenamefont {Murayama}}]{Chu:2018fzy}%
  \BibitemOpen
  \bibfield  {author} {\bibinfo {author} {\bibfnamefont {Xiaoyong}\
  \bibnamefont {Chu}}, \bibinfo {author} {\bibfnamefont {Camilo}\ \bibnamefont
  {Garcia-Cely}}, \ and\ \bibinfo {author} {\bibfnamefont {Hitoshi}\
  \bibnamefont {Murayama}},\ }\bibfield  {title} {\enquote {\bibinfo {title}
  {{Velocity Dependence from Resonant Self-Interacting Dark Matter}},}\ }\href
  {\doibase 10.1103/PhysRevLett.122.071103} {\bibfield  {journal} {\bibinfo
  {journal} {Phys. Rev. Lett.}\ }\textbf {\bibinfo {volume} {122}},\ \bibinfo
  {pages} {071103} (\bibinfo {year} {2019})},\ \Eprint
  {http://arxiv.org/abs/1810.04709} {arXiv:1810.04709 [hep-ph]} \BibitemShut
  {NoStop}%
\bibitem [{\citenamefont {Chu}\ \emph {et~al.}(2020{\natexlab{a}})\citenamefont
  {Chu}, \citenamefont {Garcia-Cely},\ and\ \citenamefont
  {Murayama}}]{Chu:2018faw}%
  \BibitemOpen
  \bibfield  {author} {\bibinfo {author} {\bibfnamefont {Xiaoyong}\
  \bibnamefont {Chu}}, \bibinfo {author} {\bibfnamefont {Camilo}\ \bibnamefont
  {Garcia-Cely}}, \ and\ \bibinfo {author} {\bibfnamefont {Hitoshi}\
  \bibnamefont {Murayama}},\ }\bibfield  {title} {\enquote {\bibinfo {title}
  {{Finite-size dark matter and its effect on small-scale structure}},}\ }\href
  {\doibase 10.1103/PhysRevLett.124.041101} {\bibfield  {journal} {\bibinfo
  {journal} {Phys. Rev. Lett.}\ }\textbf {\bibinfo {volume} {124}},\ \bibinfo
  {pages} {041101} (\bibinfo {year} {2020}{\natexlab{a}})},\ \Eprint
  {http://arxiv.org/abs/1901.00075} {arXiv:1901.00075 [hep-ph]} \BibitemShut
  {NoStop}%
\bibitem [{\citenamefont {Chu}\ \emph {et~al.}(2020{\natexlab{b}})\citenamefont
  {Chu}, \citenamefont {Garcia-Cely},\ and\ \citenamefont
  {Murayama}}]{Chu:2019awd}%
  \BibitemOpen
  \bibfield  {author} {\bibinfo {author} {\bibfnamefont {Xiaoyong}\
  \bibnamefont {Chu}}, \bibinfo {author} {\bibfnamefont {Camilo}\ \bibnamefont
  {Garcia-Cely}}, \ and\ \bibinfo {author} {\bibfnamefont {Hitoshi}\
  \bibnamefont {Murayama}},\ }\bibfield  {title} {\enquote {\bibinfo {title}
  {{A Practical and Consistent Parametrization of Dark Matter
  Self-Interactions}},}\ }\href {\doibase 10.1088/1475-7516/2020/06/043}
  {\bibfield  {journal} {\bibinfo  {journal} {JCAP}\ }\textbf {\bibinfo
  {volume} {06}},\ \bibinfo {pages} {043} (\bibinfo {year}
  {2020}{\natexlab{b}})},\ \Eprint {http://arxiv.org/abs/1908.06067}
  {arXiv:1908.06067 [hep-ph]} \BibitemShut {NoStop}%
\bibitem [{\citenamefont {Costantino}\ \emph {et~al.}(2020)\citenamefont
  {Costantino}, \citenamefont {Fichet},\ and\ \citenamefont
  {Tanedo}}]{Costantino:2019ixl}%
  \BibitemOpen
  \bibfield  {author} {\bibinfo {author} {\bibfnamefont {Alexandria}\
  \bibnamefont {Costantino}}, \bibinfo {author} {\bibfnamefont {Sylvain}\
  \bibnamefont {Fichet}}, \ and\ \bibinfo {author} {\bibfnamefont {Philip}\
  \bibnamefont {Tanedo}},\ }\bibfield  {title} {\enquote {\bibinfo {title}
  {{Exotic Spin-Dependent Forces from a Hidden Sector}},}\ }\href {\doibase
  10.1007/JHEP03(2020)148} {\bibfield  {journal} {\bibinfo  {journal} {JHEP}\
  }\textbf {\bibinfo {volume} {03}},\ \bibinfo {pages} {148} (\bibinfo {year}
  {2020})},\ \Eprint {http://arxiv.org/abs/1910.02972} {arXiv:1910.02972
  [hep-ph]} \BibitemShut {NoStop}%
\bibitem [{\citenamefont {Agrawal}\ \emph {et~al.}(2020)\citenamefont
  {Agrawal}, \citenamefont {Parikh},\ and\ \citenamefont
  {Reece}}]{Agrawal:2020lea}%
  \BibitemOpen
  \bibfield  {author} {\bibinfo {author} {\bibfnamefont {Prateek}\ \bibnamefont
  {Agrawal}}, \bibinfo {author} {\bibfnamefont {Aditya}\ \bibnamefont
  {Parikh}}, \ and\ \bibinfo {author} {\bibfnamefont {Matthew}\ \bibnamefont
  {Reece}},\ }\bibfield  {title} {\enquote {\bibinfo {title} {{Systematizing
  the Effective Theory of Self-Interacting Dark Matter}},}\ }\href {\doibase
  10.1007/JHEP10(2020)191} {\bibfield  {journal} {\bibinfo  {journal} {JHEP}\
  }\textbf {\bibinfo {volume} {10}},\ \bibinfo {pages} {191} (\bibinfo {year}
  {2020})},\ \Eprint {http://arxiv.org/abs/2003.00021} {arXiv:2003.00021
  [hep-ph]} \BibitemShut {NoStop}%
\bibitem [{\citenamefont {Tsai}\ \emph {et~al.}(2020)\citenamefont {Tsai},
  \citenamefont {McGehee},\ and\ \citenamefont {Murayama}}]{Tsai:2020vpi}%
  \BibitemOpen
  \bibfield  {author} {\bibinfo {author} {\bibfnamefont {Yu-Dai}\ \bibnamefont
  {Tsai}}, \bibinfo {author} {\bibfnamefont {Robert}\ \bibnamefont {McGehee}},
  \ and\ \bibinfo {author} {\bibfnamefont {Hitoshi}\ \bibnamefont {Murayama}},\
  }\bibfield  {title} {\enquote {\bibinfo {title} {{Resonant Self-Interacting
  Dark Matter from Dark QCD}},}\ }\href@noop {} {\  (\bibinfo {year} {2020})},\
  \Eprint {http://arxiv.org/abs/2008.08608} {arXiv:2008.08608 [hep-ph]}
  \BibitemShut {NoStop}%
\bibitem [{\citenamefont {Chaffey}\ \emph {et~al.}(2021)\citenamefont
  {Chaffey}, \citenamefont {Fichet},\ and\ \citenamefont
  {Tanedo}}]{Chaffey:2021tmj}%
  \BibitemOpen
  \bibfield  {author} {\bibinfo {author} {\bibfnamefont {Ian}\ \bibnamefont
  {Chaffey}}, \bibinfo {author} {\bibfnamefont {Sylvain}\ \bibnamefont
  {Fichet}}, \ and\ \bibinfo {author} {\bibfnamefont {Philip}\ \bibnamefont
  {Tanedo}},\ }\bibfield  {title} {\enquote {\bibinfo {title}
  {{Continuum-Mediated Self-Interacting Dark Matter}},}\ }\href {\doibase
  10.1007/JHEP06(2021)008} {\bibfield  {journal} {\bibinfo  {journal} {JHEP}\
  }\textbf {\bibinfo {volume} {06}},\ \bibinfo {pages} {008} (\bibinfo {year}
  {2021})},\ \Eprint {http://arxiv.org/abs/2102.05674} {arXiv:2102.05674
  [hep-ph]} \BibitemShut {NoStop}%
\bibitem [{\citenamefont {Kaplinghat}\ \emph {et~al.}(2016)\citenamefont
  {Kaplinghat}, \citenamefont {Tulin},\ and\ \citenamefont
  {Yu}}]{Kaplinghat:2015aga}%
  \BibitemOpen
  \bibfield  {author} {\bibinfo {author} {\bibfnamefont {Manoj}\ \bibnamefont
  {Kaplinghat}}, \bibinfo {author} {\bibfnamefont {Sean}\ \bibnamefont
  {Tulin}}, \ and\ \bibinfo {author} {\bibfnamefont {Hai-Bo}\ \bibnamefont
  {Yu}},\ }\bibfield  {title} {\enquote {\bibinfo {title} {{Dark Matter Halos
  as Particle Colliders: Unified Solution to Small-Scale Structure Puzzles from
  Dwarfs to Clusters}},}\ }\href {\doibase 10.1103/PhysRevLett.116.041302}
  {\bibfield  {journal} {\bibinfo  {journal} {Phys. Rev. Lett.}\ }\textbf
  {\bibinfo {volume} {116}},\ \bibinfo {pages} {041302} (\bibinfo {year}
  {2016})},\ \Eprint {http://arxiv.org/abs/1508.03339} {arXiv:1508.03339
  [astro-ph.CO]} \BibitemShut {NoStop}%
\bibitem [{\citenamefont {Orlofsky}\ and\ \citenamefont
  {Zhang}(2021)}]{Orlofsky:2021mmy}%
  \BibitemOpen
  \bibfield  {author} {\bibinfo {author} {\bibfnamefont {Nicholas}\
  \bibnamefont {Orlofsky}}\ and\ \bibinfo {author} {\bibfnamefont {Yue}\
  \bibnamefont {Zhang}},\ }\bibfield  {title} {\enquote {\bibinfo {title}
  {{Neutrino as the dark force}},}\ }\href {\doibase
  10.1103/PhysRevD.104.075010} {\bibfield  {journal} {\bibinfo  {journal}
  {Phys. Rev. D}\ }\textbf {\bibinfo {volume} {104}},\ \bibinfo {pages}
  {075010} (\bibinfo {year} {2021})},\ \Eprint
  {http://arxiv.org/abs/2106.08339} {arXiv:2106.08339 [hep-ph]} \BibitemShut
  {NoStop}%
\bibitem [{\citenamefont {Falkowski}\ \emph {et~al.}(2009)\citenamefont
  {Falkowski}, \citenamefont {Juknevich},\ and\ \citenamefont
  {Shelton}}]{Falkowski:2009yz}%
  \BibitemOpen
  \bibfield  {author} {\bibinfo {author} {\bibfnamefont {Adam}\ \bibnamefont
  {Falkowski}}, \bibinfo {author} {\bibfnamefont {Jose}\ \bibnamefont
  {Juknevich}}, \ and\ \bibinfo {author} {\bibfnamefont {Jessie}\ \bibnamefont
  {Shelton}},\ }\bibfield  {title} {\enquote {\bibinfo {title} {{Dark Matter
  Through the Neutrino Portal}},}\ }\href@noop {} {\  (\bibinfo {year}
  {2009})},\ \Eprint {http://arxiv.org/abs/0908.1790} {arXiv:0908.1790
  [hep-ph]} \BibitemShut {NoStop}%
\bibitem [{\citenamefont {Ko}\ and\ \citenamefont {Tang}(2014)}]{Ko:2014bka}%
  \BibitemOpen
  \bibfield  {author} {\bibinfo {author} {\bibfnamefont {P.}~\bibnamefont
  {Ko}}\ and\ \bibinfo {author} {\bibfnamefont {Yong}\ \bibnamefont {Tang}},\
  }\bibfield  {title} {\enquote {\bibinfo {title}
  {{\ensuremath{\nu}\ensuremath{\Lambda}MDM: A model for sterile neutrino and
  dark matter reconciles cosmological and neutrino oscillation data after
  BICEP2}},}\ }\href {\doibase 10.1016/j.physletb.2014.10.035} {\bibfield
  {journal} {\bibinfo  {journal} {Phys. Lett. B}\ }\textbf {\bibinfo {volume}
  {739}},\ \bibinfo {pages} {62--67} (\bibinfo {year} {2014})},\ \Eprint
  {http://arxiv.org/abs/1404.0236} {arXiv:1404.0236 [hep-ph]} \BibitemShut
  {NoStop}%
\bibitem [{\citenamefont {Bertoni}\ \emph {et~al.}(2015)\citenamefont
  {Bertoni}, \citenamefont {Ipek}, \citenamefont {McKeen},\ and\ \citenamefont
  {Nelson}}]{Bertoni:2014mva}%
  \BibitemOpen
  \bibfield  {author} {\bibinfo {author} {\bibfnamefont {Bridget}\ \bibnamefont
  {Bertoni}}, \bibinfo {author} {\bibfnamefont {Seyda}\ \bibnamefont {Ipek}},
  \bibinfo {author} {\bibfnamefont {David}\ \bibnamefont {McKeen}}, \ and\
  \bibinfo {author} {\bibfnamefont {Ann~E.}\ \bibnamefont {Nelson}},\
  }\bibfield  {title} {\enquote {\bibinfo {title} {{Constraints and
  consequences of reducing small scale structure via large dark matter-neutrino
  interactions}},}\ }\href {\doibase 10.1007/JHEP04(2015)170} {\bibfield
  {journal} {\bibinfo  {journal} {JHEP}\ }\textbf {\bibinfo {volume} {04}},\
  \bibinfo {pages} {170} (\bibinfo {year} {2015})},\ \Eprint
  {http://arxiv.org/abs/1412.3113} {arXiv:1412.3113 [hep-ph]} \BibitemShut
  {NoStop}%
\bibitem [{\citenamefont {Batell}\ \emph
  {et~al.}(2018{\natexlab{a}})\citenamefont {Batell}, \citenamefont {Han},\
  and\ \citenamefont {Shams Es~Haghi}}]{Batell:2017rol}%
  \BibitemOpen
  \bibfield  {author} {\bibinfo {author} {\bibfnamefont {Brian}\ \bibnamefont
  {Batell}}, \bibinfo {author} {\bibfnamefont {Tao}\ \bibnamefont {Han}}, \
  and\ \bibinfo {author} {\bibfnamefont {Barmak}\ \bibnamefont {Shams
  Es~Haghi}},\ }\bibfield  {title} {\enquote {\bibinfo {title} {{Indirect
  Detection of Neutrino Portal Dark Matter}},}\ }\href {\doibase
  10.1103/PhysRevD.97.095020} {\bibfield  {journal} {\bibinfo  {journal} {Phys.
  Rev. D}\ }\textbf {\bibinfo {volume} {97}},\ \bibinfo {pages} {095020}
  (\bibinfo {year} {2018}{\natexlab{a}})},\ \Eprint
  {http://arxiv.org/abs/1704.08708} {arXiv:1704.08708 [hep-ph]} \BibitemShut
  {NoStop}%
\bibitem [{\citenamefont {Berryman}\ \emph {et~al.}(2017)\citenamefont
  {Berryman}, \citenamefont {de~Gouv\^ea}, \citenamefont {Kelly},\ and\
  \citenamefont {Zhang}}]{Berryman:2017twh}%
  \BibitemOpen
  \bibfield  {author} {\bibinfo {author} {\bibfnamefont {Jeffrey~M.}\
  \bibnamefont {Berryman}}, \bibinfo {author} {\bibfnamefont {Andr\'e}\
  \bibnamefont {de~Gouv\^ea}}, \bibinfo {author} {\bibfnamefont {Kevin~J.}\
  \bibnamefont {Kelly}}, \ and\ \bibinfo {author} {\bibfnamefont {Yue}\
  \bibnamefont {Zhang}},\ }\bibfield  {title} {\enquote {\bibinfo {title}
  {{Dark Matter and Neutrino Mass from the Smallest Non-Abelian Chiral Dark
  Sector}},}\ }\href {\doibase 10.1103/PhysRevD.96.075010} {\bibfield
  {journal} {\bibinfo  {journal} {Phys. Rev. D}\ }\textbf {\bibinfo {volume}
  {96}},\ \bibinfo {pages} {075010} (\bibinfo {year} {2017})},\ \Eprint
  {http://arxiv.org/abs/1706.02722} {arXiv:1706.02722 [hep-ph]} \BibitemShut
  {NoStop}%
\bibitem [{\citenamefont {Schmaltz}\ and\ \citenamefont
  {Weiner}(2019)}]{Schmaltz:2017oov}%
  \BibitemOpen
  \bibfield  {author} {\bibinfo {author} {\bibfnamefont {Martin}\ \bibnamefont
  {Schmaltz}}\ and\ \bibinfo {author} {\bibfnamefont {Neal}\ \bibnamefont
  {Weiner}},\ }\bibfield  {title} {\enquote {\bibinfo {title} {{A Portalino to
  the Dark Sector}},}\ }\href {\doibase 10.1007/JHEP02(2019)105} {\bibfield
  {journal} {\bibinfo  {journal} {JHEP}\ }\textbf {\bibinfo {volume} {02}},\
  \bibinfo {pages} {105} (\bibinfo {year} {2019})},\ \Eprint
  {http://arxiv.org/abs/1709.09164} {arXiv:1709.09164 [hep-ph]} \BibitemShut
  {NoStop}%
\bibitem [{\citenamefont {Batell}\ \emph
  {et~al.}(2018{\natexlab{b}})\citenamefont {Batell}, \citenamefont {Han},
  \citenamefont {McKeen},\ and\ \citenamefont {Shams
  Es~Haghi}}]{Batell:2017cmf}%
  \BibitemOpen
  \bibfield  {author} {\bibinfo {author} {\bibfnamefont {Brian}\ \bibnamefont
  {Batell}}, \bibinfo {author} {\bibfnamefont {Tao}\ \bibnamefont {Han}},
  \bibinfo {author} {\bibfnamefont {David}\ \bibnamefont {McKeen}}, \ and\
  \bibinfo {author} {\bibfnamefont {Barmak}\ \bibnamefont {Shams Es~Haghi}},\
  }\bibfield  {title} {\enquote {\bibinfo {title} {{Thermal Dark Matter Through
  the Dirac Neutrino Portal}},}\ }\href {\doibase 10.1103/PhysRevD.97.075016}
  {\bibfield  {journal} {\bibinfo  {journal} {Phys. Rev. D}\ }\textbf {\bibinfo
  {volume} {97}},\ \bibinfo {pages} {075016} (\bibinfo {year}
  {2018}{\natexlab{b}})},\ \Eprint {http://arxiv.org/abs/1709.07001}
  {arXiv:1709.07001 [hep-ph]} \BibitemShut {NoStop}%
\bibitem [{\citenamefont {Becker}(2019)}]{Becker:2018rve}%
  \BibitemOpen
  \bibfield  {author} {\bibinfo {author} {\bibfnamefont {Mathias}\ \bibnamefont
  {Becker}},\ }\bibfield  {title} {\enquote {\bibinfo {title} {{Dark Matter
  from Freeze-In via the Neutrino Portal}},}\ }\href {\doibase
  10.1140/epjc/s10052-019-7095-7} {\bibfield  {journal} {\bibinfo  {journal}
  {Eur. Phys. J. C}\ }\textbf {\bibinfo {volume} {79}},\ \bibinfo {pages} {611}
  (\bibinfo {year} {2019})},\ \Eprint {http://arxiv.org/abs/1806.08579}
  {arXiv:1806.08579 [hep-ph]} \BibitemShut {NoStop}%
\bibitem [{\citenamefont {Folgado}\ \emph {et~al.}(2018)\citenamefont
  {Folgado}, \citenamefont {G\'omez-Vargas}, \citenamefont {Rius},\ and\
  \citenamefont {Ruiz De~Austri}}]{Folgado:2018qlv}%
  \BibitemOpen
  \bibfield  {author} {\bibinfo {author} {\bibfnamefont {Miguel~G.}\
  \bibnamefont {Folgado}}, \bibinfo {author} {\bibfnamefont {Germ\'an~A.}\
  \bibnamefont {G\'omez-Vargas}}, \bibinfo {author} {\bibfnamefont {Nuria}\
  \bibnamefont {Rius}}, \ and\ \bibinfo {author} {\bibfnamefont {Roberto}\
  \bibnamefont {Ruiz De~Austri}},\ }\bibfield  {title} {\enquote {\bibinfo
  {title} {{Probing the sterile neutrino portal to Dark Matter with $\gamma$
  rays}},}\ }\href {\doibase 10.1088/1475-7516/2018/08/002} {\bibfield
  {journal} {\bibinfo  {journal} {JCAP}\ }\textbf {\bibinfo {volume} {08}},\
  \bibinfo {pages} {002} (\bibinfo {year} {2018})},\ \Eprint
  {http://arxiv.org/abs/1803.08934} {arXiv:1803.08934 [hep-ph]} \BibitemShut
  {NoStop}%
\bibitem [{\citenamefont {Lamprea}\ \emph {et~al.}(2021)\citenamefont
  {Lamprea}, \citenamefont {Peinado}, \citenamefont {Smolenski},\ and\
  \citenamefont {Wudka}}]{Lamprea:2019qet}%
  \BibitemOpen
  \bibfield  {author} {\bibinfo {author} {\bibfnamefont {J.~M.}\ \bibnamefont
  {Lamprea}}, \bibinfo {author} {\bibfnamefont {E.}~\bibnamefont {Peinado}},
  \bibinfo {author} {\bibfnamefont {S.}~\bibnamefont {Smolenski}}, \ and\
  \bibinfo {author} {\bibfnamefont {J.}~\bibnamefont {Wudka}},\ }\bibfield
  {title} {\enquote {\bibinfo {title} {{Self-interacting neutrino portal dark
  matter}},}\ }\href {\doibase 10.1103/PhysRevD.103.015017} {\bibfield
  {journal} {\bibinfo  {journal} {Phys. Rev. D}\ }\textbf {\bibinfo {volume}
  {103}},\ \bibinfo {pages} {015017} (\bibinfo {year} {2021})},\ \Eprint
  {http://arxiv.org/abs/1906.02340} {arXiv:1906.02340 [hep-ph]} \BibitemShut
  {NoStop}%
\bibitem [{\citenamefont {Zhang}(2022)}]{Zhang:2020nis}%
  \BibitemOpen
  \bibfield  {author} {\bibinfo {author} {\bibfnamefont {Yue}\ \bibnamefont
  {Zhang}},\ }\bibfield  {title} {\enquote {\bibinfo {title} {{Speeding up dark
  matter with solar neutrinos}},}\ }\href {\doibase 10.1093/ptep/ptab156}
  {\bibfield  {journal} {\bibinfo  {journal} {PTEP}\ }\textbf {\bibinfo
  {volume} {2022}},\ \bibinfo {pages} {013B05} (\bibinfo {year} {2022})},\
  \Eprint {http://arxiv.org/abs/2001.00948} {arXiv:2001.00948 [hep-ph]}
  \BibitemShut {NoStop}%
\bibitem [{\citenamefont {Cheng}\ and\ \citenamefont
  {Li}(1984)}]{Cheng:1984vwu}%
  \BibitemOpen
  \bibfield  {author} {\bibinfo {author} {\bibfnamefont {Ta-Pei}\ \bibnamefont
  {Cheng}}\ and\ \bibinfo {author} {\bibfnamefont {Ling-Fong}\ \bibnamefont
  {Li}},\ }\href@noop {} {\emph {\bibinfo {title} {{Gauge Theory of Elementary
  Particle Physics}}}}\ (\bibinfo  {publisher} {Oxford University Press},\
  \bibinfo {address} {Oxford, UK},\ \bibinfo {year} {1984})\BibitemShut
  {NoStop}%
\bibitem [{\citenamefont {Dobrescu}\ and\ \citenamefont
  {Mocioiu}(2006)}]{Dobrescu:2006au}%
  \BibitemOpen
  \bibfield  {author} {\bibinfo {author} {\bibfnamefont {Bogdan~A.}\
  \bibnamefont {Dobrescu}}\ and\ \bibinfo {author} {\bibfnamefont {Irina}\
  \bibnamefont {Mocioiu}},\ }\bibfield  {title} {\enquote {\bibinfo {title}
  {{Spin-dependent macroscopic forces from new particle exchange}},}\ }\href
  {\doibase 10.1088/1126-6708/2006/11/005} {\bibfield  {journal} {\bibinfo
  {journal} {JHEP}\ }\textbf {\bibinfo {volume} {11}},\ \bibinfo {pages} {005}
  (\bibinfo {year} {2006})},\ \Eprint {http://arxiv.org/abs/hep-ph/0605342}
  {arXiv:hep-ph/0605342} \BibitemShut {NoStop}%
\bibitem [{\citenamefont {Colquhoun}\ \emph {et~al.}(2021)\citenamefont
  {Colquhoun}, \citenamefont {Heeba}, \citenamefont {Kahlhoefer}, \citenamefont
  {Sagunski},\ and\ \citenamefont {Tulin}}]{Colquhoun:2020adl}%
  \BibitemOpen
  \bibfield  {author} {\bibinfo {author} {\bibfnamefont {Brian}\ \bibnamefont
  {Colquhoun}}, \bibinfo {author} {\bibfnamefont {Saniya}\ \bibnamefont
  {Heeba}}, \bibinfo {author} {\bibfnamefont {Felix}\ \bibnamefont
  {Kahlhoefer}}, \bibinfo {author} {\bibfnamefont {Laura}\ \bibnamefont
  {Sagunski}}, \ and\ \bibinfo {author} {\bibfnamefont {Sean}\ \bibnamefont
  {Tulin}},\ }\bibfield  {title} {\enquote {\bibinfo {title} {{Semiclassical
  regime for dark matter self-interactions}},}\ }\href {\doibase
  10.1103/PhysRevD.103.035006} {\bibfield  {journal} {\bibinfo  {journal}
  {Phys. Rev. D}\ }\textbf {\bibinfo {volume} {103}},\ \bibinfo {pages}
  {035006} (\bibinfo {year} {2021})},\ \Eprint
  {http://arxiv.org/abs/2011.04679} {arXiv:2011.04679 [hep-ph]} \BibitemShut
  {NoStop}%
\bibitem [{\citenamefont {Born}(1926)}]{Born:1926yhp}%
  \BibitemOpen
  \bibfield  {author} {\bibinfo {author} {\bibfnamefont {Max}\ \bibnamefont
  {Born}},\ }\bibfield  {title} {\enquote {\bibinfo {title} {{Quantenmechanik
  der Sto\ss{}vorg\"ange}},}\ }\href {\doibase 10.1007/BF01397184} {\bibfield
  {journal} {\bibinfo  {journal} {Z. Phys.}\ }\textbf {\bibinfo {volume}
  {38}},\ \bibinfo {pages} {803--827} (\bibinfo {year} {1926})}\BibitemShut
  {NoStop}%
\bibitem [{\citenamefont {Kaplan}\ \emph {et~al.}(2009)\citenamefont {Kaplan},
  \citenamefont {Luty},\ and\ \citenamefont {Zurek}}]{Kaplan:2009ag}%
  \BibitemOpen
  \bibfield  {author} {\bibinfo {author} {\bibfnamefont {David~E.}\
  \bibnamefont {Kaplan}}, \bibinfo {author} {\bibfnamefont {Markus~A.}\
  \bibnamefont {Luty}}, \ and\ \bibinfo {author} {\bibfnamefont {Kathryn~M.}\
  \bibnamefont {Zurek}},\ }\bibfield  {title} {\enquote {\bibinfo {title}
  {{Asymmetric Dark Matter}},}\ }\href {\doibase 10.1103/PhysRevD.79.115016}
  {\bibfield  {journal} {\bibinfo  {journal} {Phys. Rev. D}\ }\textbf {\bibinfo
  {volume} {79}},\ \bibinfo {pages} {115016} (\bibinfo {year} {2009})},\
  \Eprint {http://arxiv.org/abs/0901.4117} {arXiv:0901.4117 [hep-ph]}
  \BibitemShut {NoStop}%
\bibitem [{\citenamefont {Kolb}\ and\ \citenamefont
  {Turner}(1990)}]{Kolb:1990vq}%
  \BibitemOpen
  \bibfield  {author} {\bibinfo {author} {\bibfnamefont {Edward~W.}\
  \bibnamefont {Kolb}}\ and\ \bibinfo {author} {\bibfnamefont {Michael~S.}\
  \bibnamefont {Turner}},\ }\href {\doibase 10.1201/9780429492860} {\emph
  {\bibinfo {title} {{The Early Universe}}}},\ Vol.~\bibinfo {volume} {69}\
  (\bibinfo {year} {1990})\BibitemShut {NoStop}%
\bibitem [{\citenamefont {Aprile}\ \emph {et~al.}(2023)\citenamefont {Aprile}
  \emph {et~al.}}]{XENON:2023cxc}%
  \BibitemOpen
  \bibfield  {author} {\bibinfo {author} {\bibfnamefont {E.}~\bibnamefont
  {Aprile}} \emph {et~al.} (\bibinfo {collaboration} {XENON}),\ }\bibfield
  {title} {\enquote {\bibinfo {title} {{First Dark Matter Search with Nuclear
  Recoils from the XENONnT Experiment}},}\ }\href {\doibase
  10.1103/PhysRevLett.131.041003} {\bibfield  {journal} {\bibinfo  {journal}
  {Phys. Rev. Lett.}\ }\textbf {\bibinfo {volume} {131}},\ \bibinfo {pages}
  {041003} (\bibinfo {year} {2023})},\ \Eprint
  {http://arxiv.org/abs/2303.14729} {arXiv:2303.14729 [hep-ex]} \BibitemShut
  {NoStop}%
\bibitem [{\citenamefont {Agnes}\ \emph {et~al.}(2023)\citenamefont {Agnes}
  \emph {et~al.}}]{DarkSide-50:2022qzh}%
  \BibitemOpen
  \bibfield  {author} {\bibinfo {author} {\bibfnamefont {P.}~\bibnamefont
  {Agnes}} \emph {et~al.} (\bibinfo {collaboration} {DarkSide-50}),\ }\bibfield
   {title} {\enquote {\bibinfo {title} {{Search for low-mass dark matter WIMPs
  with 12~ton-day exposure of DarkSide-50}},}\ }\href {\doibase
  10.1103/PhysRevD.107.063001} {\bibfield  {journal} {\bibinfo  {journal}
  {Phys. Rev. D}\ }\textbf {\bibinfo {volume} {107}},\ \bibinfo {pages}
  {063001} (\bibinfo {year} {2023})},\ \Eprint
  {http://arxiv.org/abs/2207.11966} {arXiv:2207.11966 [hep-ex]} \BibitemShut
  {NoStop}%
\bibitem [{\citenamefont {Heintze}\ \emph {et~al.}(1979)\citenamefont {Heintze}
  \emph {et~al.}}]{Heintze:1977kk}%
  \BibitemOpen
  \bibfield  {author} {\bibinfo {author} {\bibfnamefont {J.}~\bibnamefont
  {Heintze}} \emph {et~al.},\ }\bibfield  {title} {\enquote {\bibinfo {title}
  {{A Measurement of the K+ --\ensuremath{>} e+ Neutrino gamma Structure
  Decay}},}\ }\href {\doibase 10.1016/0550-3213(79)90001-4} {\bibfield
  {journal} {\bibinfo  {journal} {Nucl. Phys. B}\ }\textbf {\bibinfo {volume}
  {149}},\ \bibinfo {pages} {365--380} (\bibinfo {year} {1979})}\BibitemShut
  {NoStop}%
\bibitem [{\citenamefont {Cortina~Gil}\ \emph {et~al.}(2021)\citenamefont
  {Cortina~Gil} \emph {et~al.}}]{NA62:2021bji}%
  \BibitemOpen
  \bibfield  {author} {\bibinfo {author} {\bibfnamefont {Eduardo}\ \bibnamefont
  {Cortina~Gil}} \emph {et~al.} (\bibinfo {collaboration} {NA62}),\ }\bibfield
  {title} {\enquote {\bibinfo {title} {{Search for $K^+$ decays to a muon and
  invisible particles}},}\ }\href {\doibase 10.1016/j.physletb.2021.136259}
  {\bibfield  {journal} {\bibinfo  {journal} {Phys. Lett. B}\ }\textbf
  {\bibinfo {volume} {816}},\ \bibinfo {pages} {136259} (\bibinfo {year}
  {2021})},\ \Eprint {http://arxiv.org/abs/2101.12304} {arXiv:2101.12304
  [hep-ex]} \BibitemShut {NoStop}%
\bibitem [{\citenamefont {Aguilar-Arevalo}\ \emph {et~al.}(2020)\citenamefont
  {Aguilar-Arevalo} \emph {et~al.}}]{PIENU:2020las}%
  \BibitemOpen
  \bibfield  {author} {\bibinfo {author} {\bibfnamefont {A.}~\bibnamefont
  {Aguilar-Arevalo}} \emph {et~al.} (\bibinfo {collaboration} {PIENU}),\
  }\bibfield  {title} {\enquote {\bibinfo {title} {{Search for the rare decays
  $\pi^+ \to \mu^+\nu_\mu\nu\bar\nu$ and $\pi^+ \to e^+\nu_e\nu\bar\nu$}},}\
  }\href {\doibase 10.1103/PhysRevD.102.012001} {\bibfield  {journal} {\bibinfo
   {journal} {Phys. Rev. D}\ }\textbf {\bibinfo {volume} {102}},\ \bibinfo
  {pages} {012001} (\bibinfo {year} {2020})},\ \Eprint
  {http://arxiv.org/abs/2006.00389} {arXiv:2006.00389 [hep-ex]} \BibitemShut
  {NoStop}%
\bibitem [{\citenamefont {Aghanim}\ \emph {et~al.}(2020)\citenamefont {Aghanim}
  \emph {et~al.}}]{Planck:2018vyg}%
  \BibitemOpen
  \bibfield  {author} {\bibinfo {author} {\bibfnamefont {N.}~\bibnamefont
  {Aghanim}} \emph {et~al.} (\bibinfo {collaboration} {Planck}),\ }\bibfield
  {title} {\enquote {\bibinfo {title} {{Planck 2018 results. VI. Cosmological
  parameters}},}\ }\href {\doibase 10.1051/0004-6361/201833910} {\bibfield
  {journal} {\bibinfo  {journal} {Astron. Astrophys.}\ }\textbf {\bibinfo
  {volume} {641}},\ \bibinfo {pages} {A6} (\bibinfo {year} {2020})},\ \bibinfo
  {note} {[Erratum: Astron.Astrophys. 652, C4 (2021)]},\ \Eprint
  {http://arxiv.org/abs/1807.06209} {arXiv:1807.06209 [astro-ph.CO]}
  \BibitemShut {NoStop}%
\bibitem [{\citenamefont {Harvey}\ \emph {et~al.}(2015)\citenamefont {Harvey},
  \citenamefont {Massey}, \citenamefont {Kitching}, \citenamefont {Taylor},\
  and\ \citenamefont {Tittley}}]{Harvey:2015hha}%
  \BibitemOpen
  \bibfield  {author} {\bibinfo {author} {\bibfnamefont {David}\ \bibnamefont
  {Harvey}}, \bibinfo {author} {\bibfnamefont {Richard}\ \bibnamefont
  {Massey}}, \bibinfo {author} {\bibfnamefont {Thomas}\ \bibnamefont
  {Kitching}}, \bibinfo {author} {\bibfnamefont {Andy}\ \bibnamefont {Taylor}},
  \ and\ \bibinfo {author} {\bibfnamefont {Eric}\ \bibnamefont {Tittley}},\
  }\bibfield  {title} {\enquote {\bibinfo {title} {{The non-gravitational
  interactions of dark matter in colliding galaxy clusters}},}\ }\href
  {\doibase 10.1126/science.1261381} {\bibfield  {journal} {\bibinfo  {journal}
  {Science}\ }\textbf {\bibinfo {volume} {347}},\ \bibinfo {pages} {1462--1465}
  (\bibinfo {year} {2015})},\ \Eprint {http://arxiv.org/abs/1503.07675}
  {arXiv:1503.07675 [astro-ph.CO]} \BibitemShut {NoStop}%
\bibitem [{\citenamefont {Yang}\ \emph {et~al.}(2023)\citenamefont {Yang},
  \citenamefont {Nadler},\ and\ \citenamefont {Yu}}]{Yang:2022mxl}%
  \BibitemOpen
  \bibfield  {author} {\bibinfo {author} {\bibfnamefont {Daneng}\ \bibnamefont
  {Yang}}, \bibinfo {author} {\bibfnamefont {Ethan~O.}\ \bibnamefont {Nadler}},
  \ and\ \bibinfo {author} {\bibfnamefont {Hai-Bo}\ \bibnamefont {Yu}},\
  }\bibfield  {title} {\enquote {\bibinfo {title} {{Strong Dark Matter
  Self-interactions Diversify Halo Populations within and surrounding the Milky
  Way}},}\ }\href {\doibase 10.3847/1538-4357/acc73e} {\bibfield  {journal}
  {\bibinfo  {journal} {Astrophys. J.}\ }\textbf {\bibinfo {volume} {949}},\
  \bibinfo {pages} {67} (\bibinfo {year} {2023})},\ \Eprint
  {http://arxiv.org/abs/2211.13768} {arXiv:2211.13768 [astro-ph.GA]}
  \BibitemShut {NoStop}%
\end{thebibliography}%
\end{document}